# Sampling and Squeezing Electromagnetic Waves through Subwavelength Ultranarrow Regions or Openings

*Mário G. Silveirinha*[1,2] *and Nader Engheta*[1]*

(1) University of Pennsylvania, Department of Electrical and Systems Engineering, Philadelphia, PA, U.S.A., engheta@ee.upenn.edu
(2) Universidade de Coimbra, Electrical Engineering Department – Instituto de Telecomunicações, Portugal, mario.silveirinha@co.it.pt

**Abstract**

Here, we investigate the physical mechanisms that may enable squeezing a complex electromagnetic field distribution through a narrow and/or partially obstructed region with little amplitude and phase distortions. Following our recent works, such field manipulations may be made possible by a procedure in which the incoming wave is first "sampled" "pixel by pixel" using an array of metallic waveguides, and in a second step the energy corresponding to each individual pixel is "squeezed" through a very narrow channel filled with a permittivity-near zero material. In this work, we study in detail these processes in scenarios where the electromagnetic wave is compressed along a single direction of space, and present theoretical models that enable the analytical modeling of such phenomena. Full-wave results obtained with an electromagnetic simulator, demonstrate the possibility of compressing an incoming wave several folds through ultranarrow channels filled with silicon carbide. The "sampling and squeezing" concept may enable unparalleled control of electromagnetic waves in the nanoscale.

---

[*] To whom correspondence should be addressed: E-mail: engheta@ee.upenn.edu

# I. Introduction

Manipulating and guiding electromagnetic energy in the nanoscale is today one of the central problems in nano-optics, biosensing and near-field imaging [1]-[8]. Currently, most of the proposals to overcome the effects of diffraction in nanoscale-systems rely on the excitation of surface plasmon polaritons. Sometime ago [9, 10], we proposed a different strategy to tunnel electromagnetic energy through narrow channels with deeply subwavelength transverse cross-section (relatively to the direction of propagation). We demonstrated theoretically that if the narrow channel is filled with a low permittivity material with $\varepsilon$-near-zero (ENZ) the transmissivity may be greatly enhanced. Our theoretical analysis showed that by reducing the transverse dimensions of the channel it may be possible to "squeeze" more and more energy through the tight channel, independently of its total length. Some of these ideas have been experimentally demonstrated at microwaves using different methods [11]-[14].

The results reported in [9, 10] assume that the incoming wave illuminates the ENZ material along the normal direction. Only in these conditions it is possible to tunnel a significant amount of energy through the ENZ filled channel. Indeed, consistent with Snell's law, it can be shown that an oblique wave is totally reflected at the ENZ material-air interface. At first sight this may seem to restrict the possibilities of using an ENZ material to transport a complex electromagnetic field pattern (image) through a subwavelength opening/channel or around an obstacle.

In Ref. [15] we have shown how this apparent difficulty can be overcome. The fundamental idea is to first "sample" the incoming wave "pixel by pixel" using an array of metallic waveguides, and then to squeeze the energy associated with each pixel through the ultranarrow ENZ filled channel taking advantage of the "supercoupling" and insulating properties of ENZ materials [9, 10, 16]. Using such concepts, in [15] we have demonstrated the possibility of transporting and compressing a complex electromagnetic image through a tiny subwavelength hole in an opaque screen. The discretization of the electromagnetic image is achieved using a bundle of metallic nanowires. Such process is inspired in part by the works [17]-[20], which showed that an array of metallic wires may be used to transport an electromagnetic image without losing the subwavelength fine details. However, the main challenge in the problem analyzed in [15] is to bend and squeeze the nanowires through the tiny hole without losing the amplitude and phase information associated with each pixel. In fact, in general diffraction effects would imply a strong reflection at the wire bends and a strong coupling between the nanowires when the image is compressed, and this would effectively preclude the image formation. As demonstrated in [15], these problems may be avoided by embedding the nanowires in an ENZ material. In fact, the characteristically long wavelengths and the static-like response of ENZ materials may enable tunneling the sampled pixels through the tiny hole, and recuperate the image with relatively low phase and amplitude distortions.

In general, the experimental demonstration of the "sampling and squeezing" concept may be technologically challenging. However, it is possible to considerably simplify the complexity of the required device if the fields are compressed along a single direction of space. Such ideas were briefly analyzed in Ref. [15], and in particular we showed how such simplified designs may enable compressing and rerouting a complex field distribution when two microwave waveguides are connected at right angles by ultranarrow channels. In this work, we focus our attention in scenarios where the electromagnetic wave is compressed along a single direction of space, and present a detailed study of the associated physical mechanisms and an analytical model for the "sampling and squeezing" phenomenon.

This paper is organized as follows. In section II, we study the transmission properties of a set of parallel plate-waveguides. It is shown that such structure can be used to transport a nearly arbitrary field distribution between two given planes with subwavelength resolution, and effectively mimics the properties of an array of nanowires in problems in which the electromagnetic fields are essentially uniform along a given direction of space. In section III, it is formally demonstrated using analytical techniques that by embedding the metallic plates in an ENZ material, it may be possible to dramatically obstruct the propagation channel without deteriorating the transmission characteristics. In section IV, we use these concepts and full wave simulations to demonstrate the possibility of dramatically compressing the modal fields of a standard dielectric waveguide. The amplitude and phase distortions of the transmitted field are characterized. Finally, in section V, the conclusions are drawn.

In this work, the time variation of the electromagnetic fields is assumed of the form $e^{-i\omega t}$, where $\omega$ is the angular frequency.

## II. Sampling an electromagnetic wave with an array of metallic waveguides

Here, we study the propagation of electromagnetic waves through a periodic system of parallel metallic plates. Such system can be regarded as the two-dimensional analogue of a bundle of metallic wires. We will show that the array of metallic plates behaves approximately as a medium with extreme optical anisotropy [4, 22, 23], and that this property may enable the transmission of the subwavelength details of the field distribution. In particular, each metallic plate samples a specific "pixel" of the incoming wave and transports it in the form of a transverse electromagnetic (TEM) mode. Thus, the incoming wave is effectively discretized into a set of independent TEM modes. Later in the paper, we will show that this property may enable the transmission of oblique electromagnetic waves through ENZ materials, and the compression and manipulation of a complex field distribution in the subwavelength scale.

The geometry of the problem is depicted in Fig. 1a. It consists of a set of parallel plate metallic waveguides with length $L$, and normal to the $x$-direction. The thickness of the parallel plates is assumed to be negligible, and for simplicity we assume that the metal is a perfectly conducting material (PEC). The distance between adjacent plates is $a$. We suppose that the structure is uniform along the $y$-direction (the problem is intrinsically two-dimensional). The magnetic field is directed along the $y$-direction.

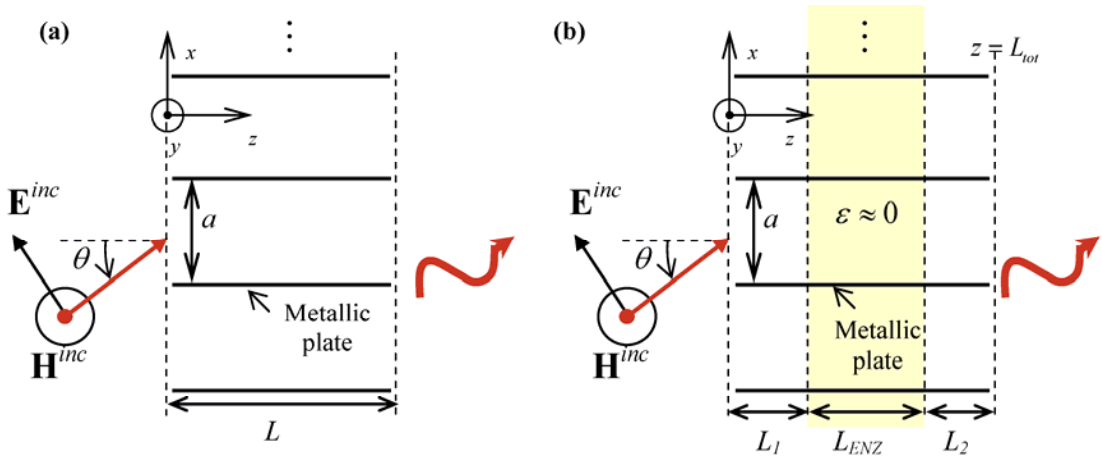

**Fig. 1.** (Color online) Geometry of the problem: (a) a plane wave impinges on a periodic set of parallel metallic plates with length $L$. The structure is uniform along the $y$-direction. (b) similar to panel (a) but the metallic plates are embedded in an ENZ slab with thickness $L_{ENZ}$.

Suppose that the set of parallel-plate waveguides is illuminated by a plane wave with $H_y^{inc} = H_0 e^{+ik_x x} e^{-\Gamma_0 z}$, where $\Gamma_0 = \sqrt{k_x^2 - (\omega/c)^2}$ is the propagation constant along $z$, and $c$ is the speed of light in vacuum. For an incoming propagating wave with $|k_x| < \omega/c$, the correct branch of the square root is such that $\Gamma_0 = -i\sqrt{(\omega/c)^2 - k_x^2}$. The angle of incidence $\theta$ is such that $k_x = (\omega/c)\sin\theta$. The magnetic field in the air regions can be written as superposition of Floquet harmonics:

$$\frac{H_y}{H_0} = e^{+ik_x x} e^{-\Gamma_0 z} + \sum_{n=-\infty}^{\infty} R_n e^{+i\left(k_x + \frac{2\pi}{a}n\right)x} e^{+\Gamma_n z}, \qquad z < 0 \tag{1a}$$

$$\frac{H_y}{H_0} = \sum_{n=-\infty}^{\infty} T_n e^{+i\left(k_x + \frac{2\pi}{a}n\right)x} e^{-\Gamma_n (z-L)}, \qquad z > L \tag{1b}$$

$$\Gamma_n = \sqrt{\left(k_x + \frac{2\pi}{a}n\right)^2 - \left(\frac{\omega}{c}\right)^2} \tag{1c}$$

where $R_n$ and $T_n$ are the complex coefficients of the reflected and transmitted waves, respectively. On the other hand, inside the parallel plate waveguide defined by $0 < x < a$ (with metallic walls at $x = 0$ and $x = a$), the magnetic field can be expanded into waveguide modes,

$$\frac{H_y}{H_0} = \sum_{m=0}^{\infty} \left(b_m^+ e^{-\gamma_m z} + b_m^- e^{\gamma_m z}\right)\cos\left(\frac{m\pi}{a}x\right), \qquad \gamma_m = \sqrt{\left(\frac{m\pi}{a}\right)^2 - \left(\frac{\omega}{c}\right)^2} \tag{2}$$

where $b_m^\pm$ are the coefficients of the expansion. Notice that the modal constants $\gamma_m$ inside the waveguide are independent of the angle of incidence of the incoming wave, i.e. they are independent of $k_x$. The solution of the problem can be obtained by imposing the continuity of $H_y$ and $E_x$ at the interfaces $z = 0$ and $z = L$, and for $0 < x < a$ (see also [25]).

For long wavelengths, all the waveguide modes are cut-off ($\gamma_m > 0$ for $m \geq 1$), except for the fundamental TEM mode with $m = 0$ for which $\gamma_0 = -i\omega/c$. This suggests that the incoming wave may be converted into a set of TEM modes. In order to quantify the amount of energy that can be transmitted through such periodic system, first we will assume single-mode propagation. In this way, we consider that to a first-order approximation $b_m^\pm = 0$, except for $m = 0$, and that $R_n = 0$ and $T_n = 0$, except for the fundamental Floquet mode with $n = 0$. Imposing the continuity of the transverse averaged field components $H_y$ and $\partial H_y / \partial z$ (note that $E_x$ is proportional to $\partial H_y / \partial z$) at the interfaces $z = 0$ and $z = L$, we easily find that the transmission coefficient is to a first-order approximation:

$$T = \frac{2\Gamma_0 \omega/c}{2(\omega/c)\Gamma_0 \cos(\omega L/c) + \left(\Gamma_0^2 - (\omega/c)^2\right)\sin(\omega L/c)} \quad (3)$$

A similar formula can be easily obtained for the reflection coefficient $R$. This approximate theory predicts that when the length of the metallic plates satisfies the Fabry-Perot condition, $\omega L/c = q\pi$, $q = 1, 2, ...$, then the transmission coefficient becomes $T = \mp 1$, independent of the angle of incidence, or more generally independent of $k_x$. This means that in these circumstances the set of metallic plates behaves as an ideal imaging system with infinite resolution. This effect is directly related to the "canalization" phenomenon identified in [21], and is a consequence of the extreme optical anisotropy of the artificial material formed by the parallel metallic plates. In fact, for long wavelengths such structured material can be described to a first approximation by the effective parameters $\varepsilon_{xx} = 1$ and $\varepsilon_{zz} = \infty$ (see Appendix A). Materials with extreme anisotropy have remarkable properties [23, 24]. It is known [22, 26] that a slab of an anisotropic medium with $\varepsilon_{zz} = \infty$ may enable near field image transport with subwavelength resolution. In particular, such structure may support a collective Fabry-Perot resonance that enables the anomalous transmission of all spatial harmonics, independent of $k_x$. As discussed in [21], such a property enables the transmission of not only propagating waves, but also of the evanescent spectrum associated with the subwavelength details of an image. This is also consistent with our intuition that each pair of parallel plates is essentially a waveguide that samples and transports the incoming radiation from the object plane to the observation plane in the form of a TEM wave. Intuitively, one can expect that the spatial resolution of such system is approximately *a*, independent of the wavelength. In Ref. [28], it was reported that a nanoscale waveguide array related to the one we just discussed above also enables subwavelength resolution at optical frequencies when the plasmonic properties of metals become predominant (the theory of [28] relies on the excitation of surface plasmon polaritons).

It is important to be aware that Eq. (3) is only a rough estimation of the actual electromagnetic response of the set of metallic plates. To have a more precise idea of the amount of energy that may be effectively be transmitted through such periodic system, in Fig. 2 and Fig. 3 we plot the transmission and reflection coefficients as a function of $k_x c/\omega$, for different values of $L$. It is assumed that the spacing between the plates *a* is such that $a = 0.03\lambda_0$, where $\lambda_0$ is the free-space wavelength. Note that if $|k_x c/\omega| < 1$ the incident wave is a propagating wave, whereas if $|k_x c/\omega| > 1$ it is an evanescent wave. In Fig. 2 the solid red lines represent the results obtained with (3), and the solid black lines represent the "exact" results obtained by solving the electromagnetic problem numerically by using a mode matching technique similar to the one reported in [25]. For all the examples, it is seen that for an incoming propagating wave the amplitude of the transmission coefficient is practically identical to unity. This confirms that the set of metallic plates can efficiently transform an oblique propagating wave into a set of TEM modes, and vice versa.

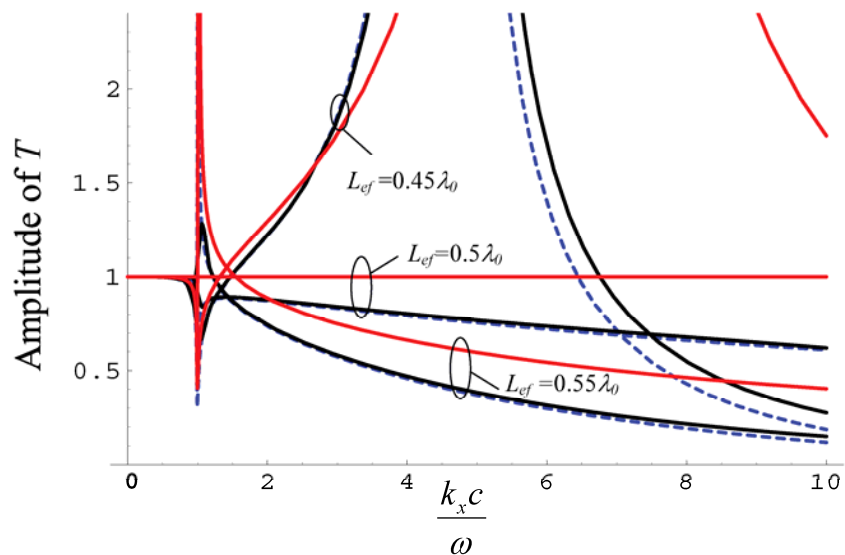

**Fig. 2.** (Color online) Amplitude of the transmission coefficient as a function of the (normalized) transverse wave vector $k_x$. The spacing between the plates is $a = 0.03\lambda_0$. The (effective) length of the plates $L_{ef}$ is indicated explicitly in the graphic for each curve. (a) Solid red line (light gray in grayscale) – approximate result obtained using single-mode theory, i.e. Eq. (3). (b) Solid black line – "exact" result obtained from the full wave numerical solution of the problem using a mode matching technique (the transmission coefficient is referred to the virtual interface: see the text for more details). (c) Blue dashed line (dark gray in grayscale) – results obtained with the spatially dispersive effective permittivity model proposed in Appendix A.

On the other hand, for evanescent waves $T$ may depend appreciably on the considered $L$. Notice that for evanescent waves, $T$ may be greater than unity without violating the conservation of energy. In particular, Fig. 2 shows that for the case $L = 0.45\lambda_0$ the transmission coefficient has a pole. This singularity corresponds to a surface wave that propagates along $x$, closely confined to the interfaces $z = 0$ and $z = L$. A similar effect has been reported in [18, 27] for a material formed by metallic wires. It is also seen that for $L = 0.5\lambda_0$ the transmission coefficient satisfies $|T| > 0.7$ for $|k_x c/\omega| < 7$, and thus the resolution of an image transported by the array of waveguides (using a half-power criterion) for these parameters is around $\lambda_0/(7 \times 2)$ (7 times better than conventional diffraction limited systems).

It is evident from Fig. 2 and Fig. 3 that the agreement between Eq. (3) and the exact solution may be relatively coarse, especially for evanescent waves. This is due to the effect of the higher order waveguide modes, which was neglected in the derivation of (3). In particular, from Fig. 2 it is seen that when the Fabry-Perot condition is satisfied, $L = 0.5\lambda_0$, the transmission coefficient is not exactly identical to unity, as predicted by Eq. (3). This confirms that the characterization of the structured material using the parameters $\varepsilon_{xx} = 1$ and $\varepsilon_{zz} = \infty$ is only a rough approximation. Fortunately, it is possible to obtain a more rigorous and accurate analytical model for the artificial material. In Appendix A, we propose to describe the structured material using the spatially dispersive permittivity model (A4). Using this nonlocal dielectric function and techniques similar to those employed in [18, 27], we obtain the linear system (A8), which can be numerically solved for the unknown $T$ and $R$. The corresponding solution is shown in Fig. 2 and Fig. 3 with a dashed blue line. It is seen that the agreement with the full wave numerical solution (solid black line) is remarkable. Similar results are obtained for other geometries, and for the phase of $T$ and $R$, provided $\omega a/c \ll \pi$ (long wavelength limit condition).

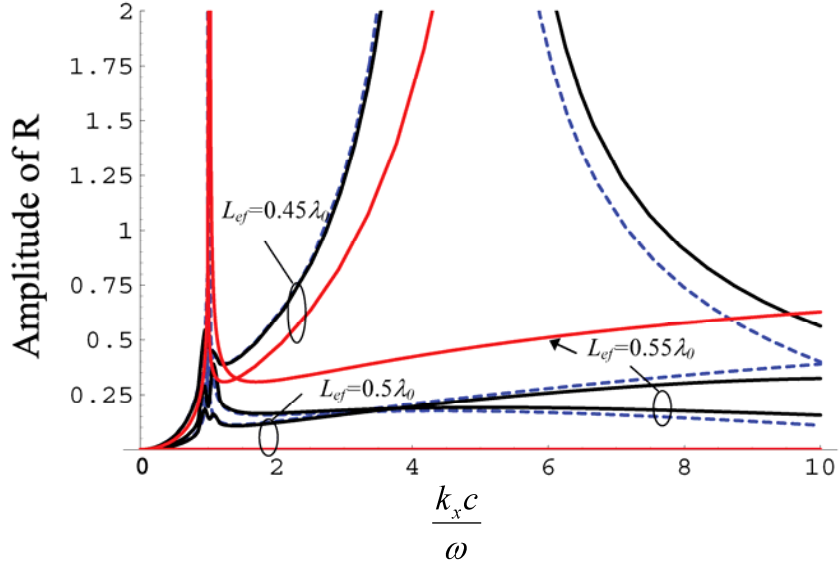

**Fig. 3.** (Color online) Amplitude of the reflection coefficient as a function of the (normalized) transverse wave vector $k_x$. The legend is as in Fig. 2.

The analytical theory developed in Appendix A is based not only on the spatially dispersive dielectric function (A4), but also on the concept of virtual interfaces introduced in [25]. In [25], we demonstrated that in order to characterize the artificial material using an effective permittivity model, it may be necessary to consider that the interfaces of the "homogenized" slab are displaced a distance $\delta$ from the actual physical interfaces. This property is a consequence of the fluctuations of the fields near the interfaces with air. Thus, a periodic set of metallic plates with length $L$ behaves effectively as an artificial material with dielectric function given by (A4), and effective length $L_{ef} = L + 2\delta$. The distance between the physical interface and the virtual interface is given by:

$$\delta \approx 0.1a \qquad (4)$$

In the long wavelength limit when $a \ll \lambda_0$, the correction provided by the virtual interfaces is in general relatively small. In the simulations shown in Fig. 2 and Fig. 3, $L_{ef}$ represents the thickness of the homogenized slab. This means that the results corresponding to the solid black curves in Fig. 2 and Fig. 3 are associated with a structure formed by a periodic set of parallel metallic plates with length given by $L = L_{ef} - 2\delta$, whereas the results corresponding to the blue and red curves are obtained by replacing $L = L_{ef}$ in the analytical models. Moreover, for the black curves the calculated transmission and reflection coefficients are referred to reference planes coincident with the virtual interfaces. These concepts are explained in detail in Appendix A.

Let us summarize the results presented in this section. It was shown that if the metallic plates are closely packed ($\omega a / c \ll \pi$), and the length of the metallic plates $L$ is tuned so that,

$$L_{ef} = q \frac{\lambda_0}{2}, \qquad q = 1, 2, .... \qquad \text{with } L = L_{ef} - 2\delta \qquad (5)$$

then the transmission coefficient $T$ is to a good approximation $T(k_x) \approx \mp 1$, for all the propagating harmonics and for a significant part of the evanescent spectrum. This property demonstrates that the structured material may enable imaging with super-resolution, using a mechanism similar to that studied in [17, 27]. Moreover, even when the length of the plates does not verify the Fabry-Perot condition (5), the approximate identity $|T(k_x)| \approx 1$ may still be valid for a relatively wide range of spatial harmonics. In particular, the set of parallel metallic plates is able of converting a nearly arbitrary field distribution into a set of TEM waves. The numerical simulations demonstrate that these phenomena may be described almost exactly by the analytical theory developed in Appendix A.

# III. Tunneling oblique waves through ENZ channels

In our previous work [9, 10], it was theoretically demonstrated that ENZ materials may enhance the transmission of electromagnetic waves through very narrow channels with bends, and that this property may have interesting potentials in several problems. In particular, it was proven that in the $\varepsilon = 0$ limit the scattering parameters are independent of the specific geometry of the channel, and are only influenced by the total area of the longitudinal cross-section of the channel. Such ENZ materials may be directly available in nature at infrared and optical frequencies when some noble metals, semiconductors, or polar dielectrics (e.g. silicon carbide) are near their plasma frequencies [30, 31, 32], or they can be engineered as metamaterials at desired frequencies.

An important feature of the tunneling phenomenon reported in [9, 10] is that it requires that the incident wave propagates along the normal to the ENZ interface. Otherwise, for oblique incidence, the wave is completely reflected. Indeed, as verified in [29], in the $\varepsilon = 0$ lossless limit an ENZ slab behaves as an angular filter with an anomalous discontinuity of the transmission coefficient: for normal incidence a wave may tunnel through the slab, while for oblique incidence the wave is completely reflected and the ENZ material behaves as a perfect magnetic conductor (PMC). For $\varepsilon$-near zero, but not identical to zero, the angular "bandwidth" of the ENZ slab is finite but extremely narrow. This is illustrated in Fig. 4 (red solid line), where we plot the transmission coefficient when a plane wave impinges on an ENZ slab with thickness $L_{ENZ} = 0.25\lambda_0$. At the considered frequency the permittivity of the ENZ material is $\varepsilon = 0.001\varepsilon_0$. As seen in Fig. 4, the transmission coefficient is nearly identical to zero for every value of $k_x$, except for a very narrow interval near $k_x \approx 0$ (paraxial incidence). Remember that for propagating waves the angle of incidence is such that $\sin\theta = k_x c / \omega$.

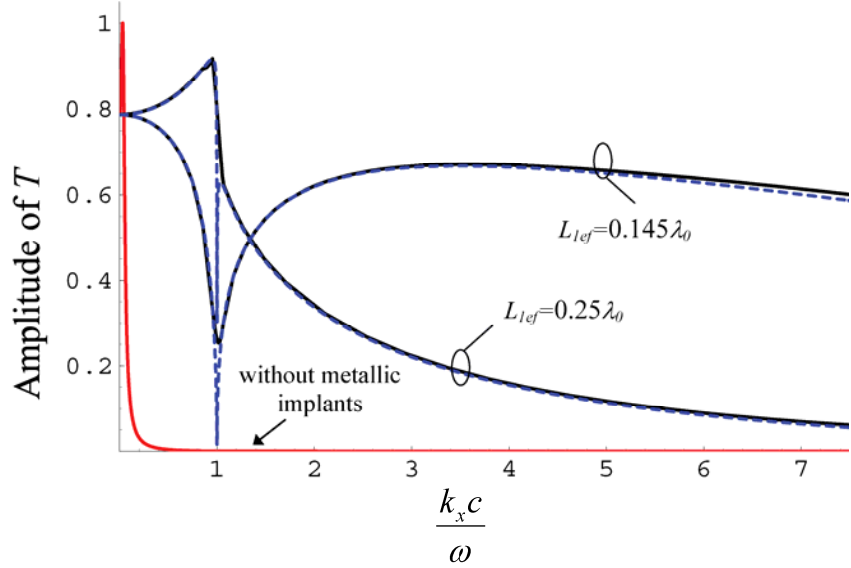

**Fig. 4.** (Color online) Amplitude of the transmission coefficient as a function of the (normalized) transverse wave vector $k_x$. The ENZ slab has permittivity $\varepsilon = 0.001$ at the frequency of interest, and normalized thickness $L_{ENZ} = 0.25\lambda_0$. (a) Solid red line (light gray in grayscale) – unloaded ENZ slab, i.e. without metallic implants. (b) Solid black line – "exact" transmission characteristic for an ENZ slab loaded with metallic plates such that $L_{1ef} = L_{2ef}$; this curve was computed using the mode-matching numerical method. The spacing between the metallic plates is such that $a = 0.03\lambda_0$. (c) Blue dashed line (dark gray in grayscale) – similar to (b) but obtained with the approximate analytical model derived in Appendix B.

The transmission characteristics of the system change dramatically when a periodic array of parallel metallic plates are embedded into the ENZ slab (Fig. 1b). Notice that the metallic plates are extended into the air regions, with lengths $L_1$ and $L_2$. As discussed in the previous section, the metallic plates may enable sampling of the incoming wave "pixel" by "pixel", transforming it into a collection of TEM modes. Since, the reflectivity of an ENZ material is minimized near the normal direction [29], it is expected that this geometry may significantly enhance the transmission for oblique incidence. This is confirmed in Fig. 4 (solid black lines) where we plot the transmission coefficient for different values of $L_1 = L_2$, supposing that the spacing between the plates is such that $a = 0.03\lambda_0$. These results were obtained using the mode matching method and can be considered "numerically exact". As in the previous section, the amplitudes of the incident and transmitted waves are referred to the virtual interfaces of the associated homogenized structure, i.e. to a distance of $\delta = 0.1a$ away from the physical interfaces (this was done so that the full wave results can be compared with the analytical theory discussed later). By definition the parameters $L_{1ef}$ and $L_{2ef}$ are such that $L_{1ef} = L_1 + \delta$ and $L_{2ef} = L_2 + \delta$.

In particular, it is seen in Fig. 4 that when $L_{1ef} = 0.25\lambda_0$ every incoming propagating plane wave ($|k_x c/\omega| < 1$) can tunnel through the system. Remarkably, the transmission coefficient even increases for oblique incidence, and has a maximum for grazing incidence ($k_x c/\omega = 1$). This behavior contrasts markedly with the case in which the ENZ slab is unloaded (solid red line). Moreover, a significant portion of the evanescent spectrum (with $|k_x c/\omega| > 1$) can also be tunneled through the ENZ slab. This demonstrates how the presence of the metallic plates completely modifies the transfer function of the system. In fact, while an ENZ slab on its own behaves as a very narrow band angular filter, an ENZ slab with metallic implants may behave as an all-pass filter, with subwavelength resolution.

As in section II, the transmission and reflection properties of the considered system can be predicted with excellent accuracy using homogenization concepts (dashed blue lines in Fig. 4). Briefly, the idea is to consider that inside each parallel-plate waveguide the electromagnetic field is the superposition of the $m=0$ and $m=1$ waveguide modes, while in the air region the wave consists only of the fundamental Floquet mode. The interaction between the fields at the interfaces with air is treated in the same way as in Appendix A. On the other hand, at the interfaces with the ENZ material, the *m*=1 waveguide mode is completely reflected, since the ENZ material behaves as a perfect magnetic conductor for oblique incidence (in the $\varepsilon = 0$ limit). Thus, the ENZ transitions can be completely characterized by the reflection coefficient $\rho_{ENZ}$ for TEM incidence (*m*=0 mode), and by the corresponding transmission coefficient $\zeta_{ENZ} = 1 + \rho_{ENZ}$. Using these concepts it is possible to reduce the scattering problem to an 8×8 linear system given by (B5), which can be solved numerically. A detailed description of the homogenization procedure can be found in Appendix B.

It is important to mention that the transmission characteristics of the system depends on the specific value of $L_1$. This is consistent with the results of the previous section, where it was found that the coupling of the incoming wave to the TEM modes was enhanced for certain specific lengths of the metallic plates. For example, it is seen in Fig. 4 that for $L_{1ef} = 0.145\lambda_0$ the transmission of evanescent waves is improved as compared to the previous example. However, the transmission of propagating oblique waves is somewhat deteriorated.

To understand how the value of $L_1 = L_2$ affects the transmission through the system, we have calculated *T* as a function of the (normalized) length $L_1$ of the plate extensions, and for different angles of incidence. The results reported in Fig. 5 show that for wide incident angles *T* varies significantly with $L_1$. Very interestingly, it can also be seen that for certain values of $L_1$ the transmission coefficient is close to unity for the entire propagating spectrum, i.e. for arbitrary $\theta$. This happens when $2L_{1,ef} \approx 0.5\lambda_0$ or $2L_{1,ef} \approx 1.5\lambda_0$ (first and third peaks of the $\theta = 80°$ transmission characteristic), or alternatively when $2L_{1,ef} \approx 0.8\lambda_0$ or $2L_{1,ef} \approx 1.8\lambda_0$ (second and fourth peaks).

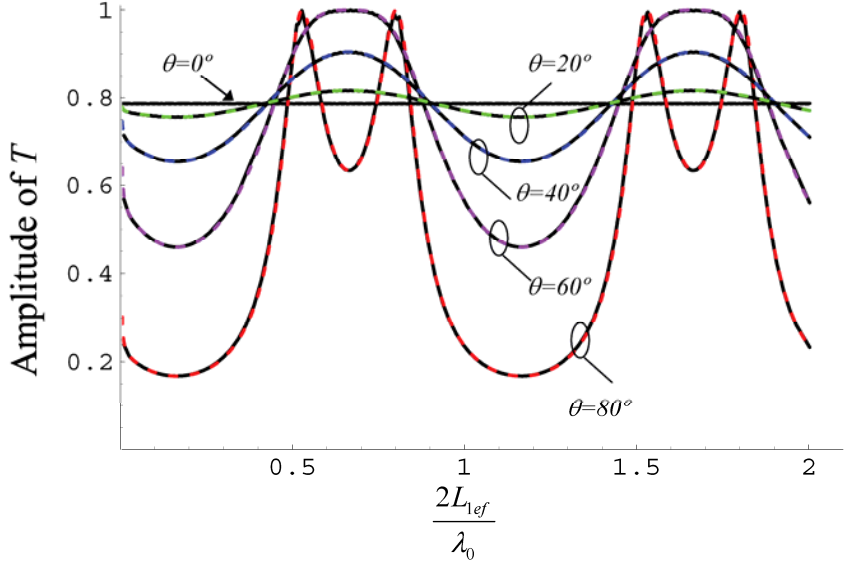

**Fig. 5.** (Color online) Amplitude of the transmission coefficient as a function of the (normalized) thickness $L_{1ef} = L_{2ef}$ of the metallic plate extensions for different angles of incidence $\theta$. The ENZ slab has permittivity $\varepsilon = 0.001$ at the frequency of interest, and normalized thickness $L_{ENZ} = 0.25\lambda_0$. The spacing between the metallic plates is such that $a = 0.03\lambda_0$. (a) Solid black lines – "exact" full wave result obtained with the mode-matching method. (b) Dashed colored (gray) lines – results obtained with the analytical model derived in Appendix B. These curves are practically coincident with the results obtained with the full wave method.

Moreover, other simulations (not shown here) demonstrate that the position of the first and third peaks is stable and nearly independent of the specific value of $L_{ENZ}$, whereas the position of the second and fourth peaks tends to decrease as $L_{ENZ}$ increases. Notice that in the $L_{ENZ} = 0$ case the considered structure reduces to the geometry studied in section II, and thus, from (5), we expect that the transmission peaks occur when $L_{1ef} + L_{2ef} \approx 0.5q\lambda_0$, $q = 1, 2, .....$. This suggests that the peaks reported in Fig. 5 are also somehow related with the Fabry-Perot resonances identified in section II.

In fact, let us consider the "odd numbered" resonances for which,

$$L_{1ef} + L_{2ef} = q\frac{\lambda_0}{2}, \quad \text{with } L_{1ef} = L_{2ef} \text{ and } q = 1, 3, 5,... \quad (6)$$

Using transmission line theory it is simple to understand why in such case the transmission properties of the system are enhanced for oblique waves. To this end, we define the characteristic transverse impedance of a propagating mode as $Z_c \equiv E_x / H_y$. In the air region the characteristic impedance is $Z_{c,air} = Z_0 \cos\theta$, where $Z_0 = \sqrt{\mu_0/\varepsilon_0}$ is the free-space impedance. On the other hand, the transverse impedance seen by the TEM mode is $Z_{c,TEM} = Z_0/\sqrt{\varepsilon_r}$, where $\varepsilon_r$ is the relative permittivity of the material in between the metallic plates ($\varepsilon_r = 1$ in the sections with length $L_1$ and $L_2$, and $\varepsilon_{r,ENZ} \approx 0$ in the section filled with the ENZ material). Next, we notice that when (6) is satisfied the waveguide sections with lengths $L_1$ and $L_2$ behave as $\lambda/4$ transformers, independent of the incident angle. This means that the system is matched when the incident angle satisfies $Z_0^2 = Z_{c,air} Z_{c,TEM}^{(ENZ)}$, or equivalently when $\cos\theta = \sqrt{\varepsilon_{r,ENZ}}$. Hence, the previous arguments demonstrate the system becomes matched for some $\theta \approx 90°$ (grazing incidence). Notice that the previous analysis is valid independent of the value of $L_{ENZ}$. This justifies that the position of the odd numbered Fabry-Perot resonances is stable with respect to variations of $L_{ENZ}$.

Since the transmission along the normal direction ($\theta = 0$) is independent of $L_1$ and $L_2$ (Fig. 5), we conclude that in order to enable the tunneling of the entire spectrum of propagating waves it is sufficient to tune $L_1$ and $L_2$ so that they satisfy (6). We underline that this choice can be made independent of the value of $L_{ENZ}$ (quite differently, the "even numbered Fabry-Perot resonances" depend significantly on $L_{ENZ}$). This property will be very useful in section IV, where $L_{ENZ}$ depends on the specific waveguide.

From a different and more general perspective, which also applies to the system of nanowires considered in Ref. [15], these results can be also justified by noting that when the length of a metallic plate verifies $L = 0.5\lambda q$, with $q=1, 3, 5, \ldots$ the electric current at the center of the metallic plate is maximal [33]. This means that if the plate is split into two identical pieces (with lengths $L_1 = L_2$), and the two pieces are connected by another metallic plate embedded in an ENZ material, the current injected in the ENZ material may be quite large (notice that due to the slow phase variation in the ENZ material, it seems plausible that the resonance condition for the current is invariant). As demonstrated in Refs. [10, 15], the ENZ material behaves as perfect insulator for the conduction current, in the sense that the current injected at the input interface is exactly reproduced at the output interface. Thus, it is clear that when Eq. (6) is satisfied the sensed current at the output interface (associated with a given pixel) may be quite significant, which explains the good transmission properties of the system in such conditions. Notice that the same reasoning also shows that when $L_1 + L_2 = 0.5\lambda_0 q$ with $q$ even, the current injected in the ENZ material is expected to be relatively small [33], which elucidates the different behavior of odd and even Fabry-Perot resonances.

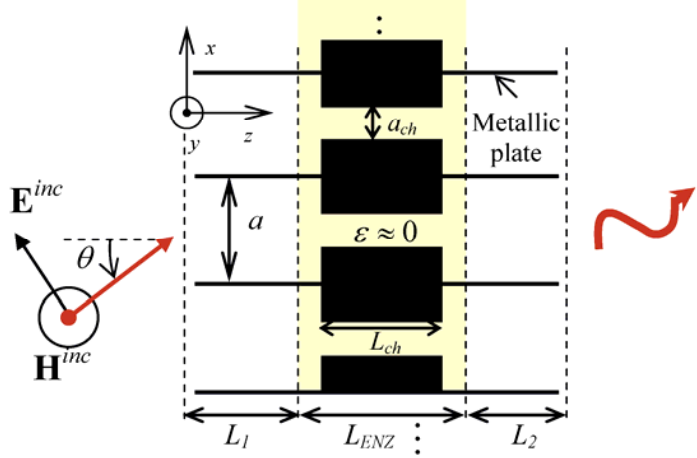

**Fig. 6.** (Color online) A plane wave impinges on a periodic array of parallel metallic plates partially embedded in an ENZ material. Inside the ENZ material the channel may be made extremely narrow, as illustrated in the figure (the dark regions represent perfect electric conductors). The structure is uniform along the *y*-direction.

The fact that the wave inside the parallel plate waveguides is essentially a TEM wave has a remarkable implication. Indeed, the results of [9, 10] show that in the $\varepsilon = 0$ limit, the reflection coefficient (under TEM incidence) in a metallic waveguide with an ENZ channel of arbitrary geometry is given by,

$$\rho_{ENZ} = \frac{(a_1 - a_2) + i\frac{\omega}{c}\mu_{r,p}A_p}{(a_1 + a_2) - i\frac{\omega}{c}\mu_{r,p}A_p}, \qquad \text{(ENZ lossless limit)} \qquad (7)$$

where $a_1$ and $a_2$ define the distance between the parallel plates at the input and output sections, $\mu_{r,p}$ is the relative permeability of the ENZ material, and $A_p$ is the total area of the cross-section of the ENZ-channel relatively to the *y*-direction (here we assume that the structure is invariant to translations along *y*). In particular, when $a \equiv a_1 = a_2$ (as in the waveguides depicted in Fig. 6) the transmission coefficient $\zeta_{ENZ} = 1 + \rho_{ENZ}$ can be made arbitrarily close to unity if $\frac{\omega}{c}\mu_{r,p}\frac{A_p}{a} \ll 1$. Hence, this implies that for the waveguides depicted in Fig. 6, the transmission can be significantly enhanced if $a_{ch}/a \to 0$ (at least if the effect of losses is small). This property suggests that the transmission of a plane wave through the periodic set of waveguides may be drastically improved if the transverse section of the ENZ channel (relatively to the propagation direction) is made tighter and tighter.

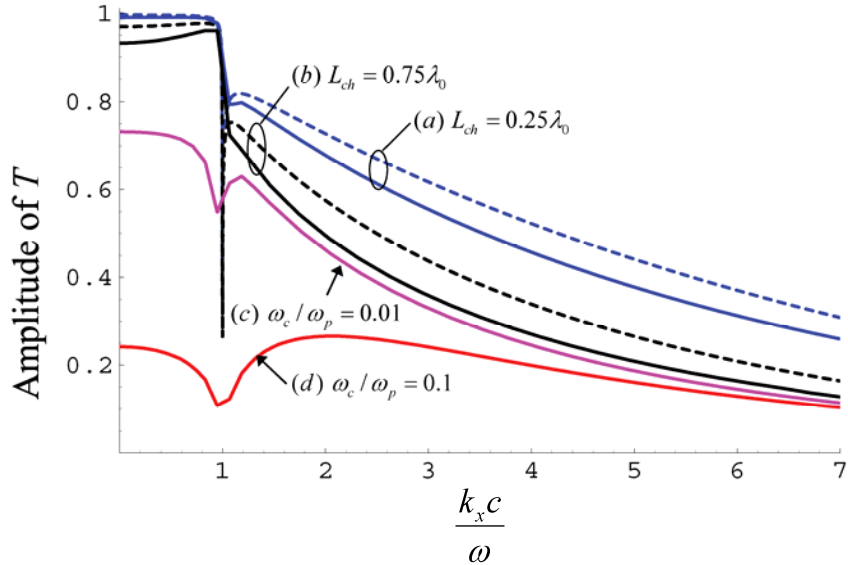

**Fig. 7.** (Color online) Amplitude of the transmission coefficient as a function of the (normalized) transverse wave vector $k_x$. The geometry of the problem is shown in Fig. 6. It is assumed that $L_{1ef} = L_{2ef} = 0.25\lambda_0$, and the spacing between the plates is such that $a = 0.03\lambda_0$. The U-shaped channel is characterized by $L_{ENZ} - L_{ch} = 0.2a$ and $a_{ch} = 0.1a$. **(a)** Length of the channel is $L_{ch} = 0.25\lambda_0$ and the losses in the ENZ material are negligible (solid line: full wave numerical results; dashed line: analytical model). **(b)** Similar to (a) but with $L_{ch} = 0.75\lambda_0$. **(c)** and **(d)** Similar to (b) but the ENZ material is characterized by absorption loss such that $\varepsilon \approx i\omega_c/\omega_p$.

In order to verify this possibility, we have used CST Microwave Studio$^{TM}$ [34] to compute the transmission characteristic for the structure depicted in Fig. 6. We considered that $a = 0.03\lambda_0$, $L_{1ef} = L_{2ef} = 0.25\lambda_0$ (where, as before, $L_{1ef} = L_1 + 0.1a$, etc) and that $L_{ENZ} = 0.2a + L_{ch}$ with $L_{ch} = 0.25\lambda_0$. Hence the geometrical parameters of the new structure are almost identical to those of the example reported in Fig. 4, except that now the ENZ channel is extremely tight with $a_{ch} = 0.1a$. The calculated transmission characteristic is depicted in Fig. 7 (curve (a)). It can be seen that the transmission coefficient is such that $|T| \approx 1$ for the entire propagating spectrum with $|k_x c/\omega| \leq 1$. This contrasts markedly with the results of Fig. 4 (solid black line associated with $L_{1ef} = 0.25\lambda_0$), where the transmission for normal incidence is notoriously lower than that for grazing incidence. This confirms our intuition that by compressing the cross-section of the ENZ channel it may be possible to significantly enhance the transmissivity and squeeze the incoming oblique plane wave through the tight channels. Moreover, such procedure also allows tunneling waves through longer channels. This is illustrated in curve (b) of Fig. 7 where we increased the length of the ENZ channel to $L_{ch} = 0.75\lambda_0$. It is remarkable that even for this long and tight channel the transmission coefficient is larger than for the geometry considered in Fig. 4. As explained in Ref. [15, 35], this supercoupling effect can also be explained in terms of impedance matching, more specifically ENZ materials may help matching the characteristic line impedances of two transmission lines with a huge height mismatch, e.g. wide and narrow parallel-plate channels (the line impedance should not be confused with the wave impedance $\sqrt{\mu/\varepsilon}$ which is always strongly mismatched in the two channels).

In the previous examples the effect of losses in the ENZ material was neglected. In order to evaluate quantitatively how this may affect the properties of the system, we consider that the ENZ material follows a Drude type dispersion model with plasma frequency $\omega_p$ and collision frequency $\omega_c$. At the plasma frequency the permittivity of the ENZ material becomes $\varepsilon \approx i\omega_c/\omega_p$. In curves (c) and (d) of Fig. 7, we plot the transmission characteristic of the channel with $L_{ch} = 0.75\lambda_0$ for the cases $\omega_c/\omega_p = 0.01$ and $\omega_c/\omega_p = 0.1$. As could be expected, the transmission coefficient drops. Nevertheless, considering that the ENZ channel is almost one-wavelength thick and is extremely narrow, one can say that the amplitude level of the transmitted wave is still acceptable. Notice that the value $\varepsilon = 0.1i$ is the value of the permittivity of silicon carbide (SiC) at its plasma frequency ($\lambda_0 = 10.31 \mu m$) [32]. In the next section, we will discuss how the proposed "sampling and squeezing" procedure may be employed to propagate a complex field distribution through a very narrow sub-wavelength channel.

It is also interesting to note that despite the apparent complexity of the geometry depicted in Fig. 6, the analytical model described in Appendix B can still predict the scattering parameters with excellent accuracy. Indeed, as mentioned before, the interaction of the waveguide modes with the ENZ transition is completely described by the reflection coefficient $\rho_{ENZ}$ for TEM incidence (*m*=0 mode), and by the corresponding transmission coefficient $\zeta_{ENZ} = 1 + \rho_{ENZ}$. But from [9] the parameter $\rho_{ENZ}$ is known in closed analytical form [see Eq. (7)], independent of the specific geometry the ENZ channel. Thus, all the arguments of the Appendix B remain valid if we consider that in formula (B5c) $A_p$ stands for the area of the longitudinal cross-section of a generic ENZ channel (in the *x-z*-plane). In particular, the scattering coefficients can still be calculated by solving the 8×8 linear system (B5). Notice that the previous discussion is valid independent of the specific geometry of the ENZ channels (which can be as depicted in Fig. 6, or truly anything else). In order to confirm these claims we have computed the transmission coefficient using the proposed analytical procedure. The results are depicted in Fig. 7 (dashed lines) showing a remarkable agreement with the full wave results obtained with CST Microwave Studio [34].

In particular, the previous arguments also show that it is not essential that all the ENZ channels are identical: they hint that the scattering parameters may be nearly independent of the specific geometry of the ENZ channels, provided all the channels have the same $\rho_{ENZ}$. It is obvious that the condition that $\rho_{ENZ}$ is invariant only requires that $A_p$ (longitudinal cross-section of a generic ENZ channel) is invariant, while the specific geometry of each channel may evidently vary. These unusual properties will be exploited in the next section.

## IV. Compressing the modal fields

The results of section III demonstrate that by using ENZ materials it may be possible to squeeze an oblique or evanescent plane wave through an array of microstructured waveguides with tight cross-sections in the *x-z* plane. This and the findings of Ref. [15] suggest that ENZ materials may be used to propagate electromagnetic energy with subwavelength mode sizes, or to couple energy between two waveguides through a very narrow aperture or through a partially obstructed path. In order to explore these possibilities, here we consider a standard (open) dielectric waveguide with thickness $h_s$ (see Fig. 8a) [36, pp. 712]. This dielectric slab has permittivity $\varepsilon_s = 2.2$ and stands in free-space ($\varepsilon_{air} = 1$). As in the previous examples, it is assumed that the structure is invariant to translations along *y*. Such approximation is sufficient to reproduce all the relevant physics in a realistic structure where the width along *y* is much larger than $h_s$. We suppose that the fundamental waveguide mode (which has no lower frequency cut-off) is excited and propagates along the *z*-direction (surface wave). The wavelength in the dielectric slab is chosen such that $\lambda_d = 1.15 h_s$ (note that $\lambda_d = \lambda_0 / \sqrt{\varepsilon_s}$). This ensures that the modal fields are essentially confined inside the dielectric slab, and decay evanescently fast in the air region. The objective here is to demonstrate the viability of squeezing the modal field several folds through a very tight subwavelength region with relatively low amplitude and phase distortion. To this end, we apply the "sampling and squeezing" concepts originally introduced in Ref. [15], and that were studied in detail in the previous sections for scenarios where the electromagnetic wave is nearly uniform along a certain direction of space.

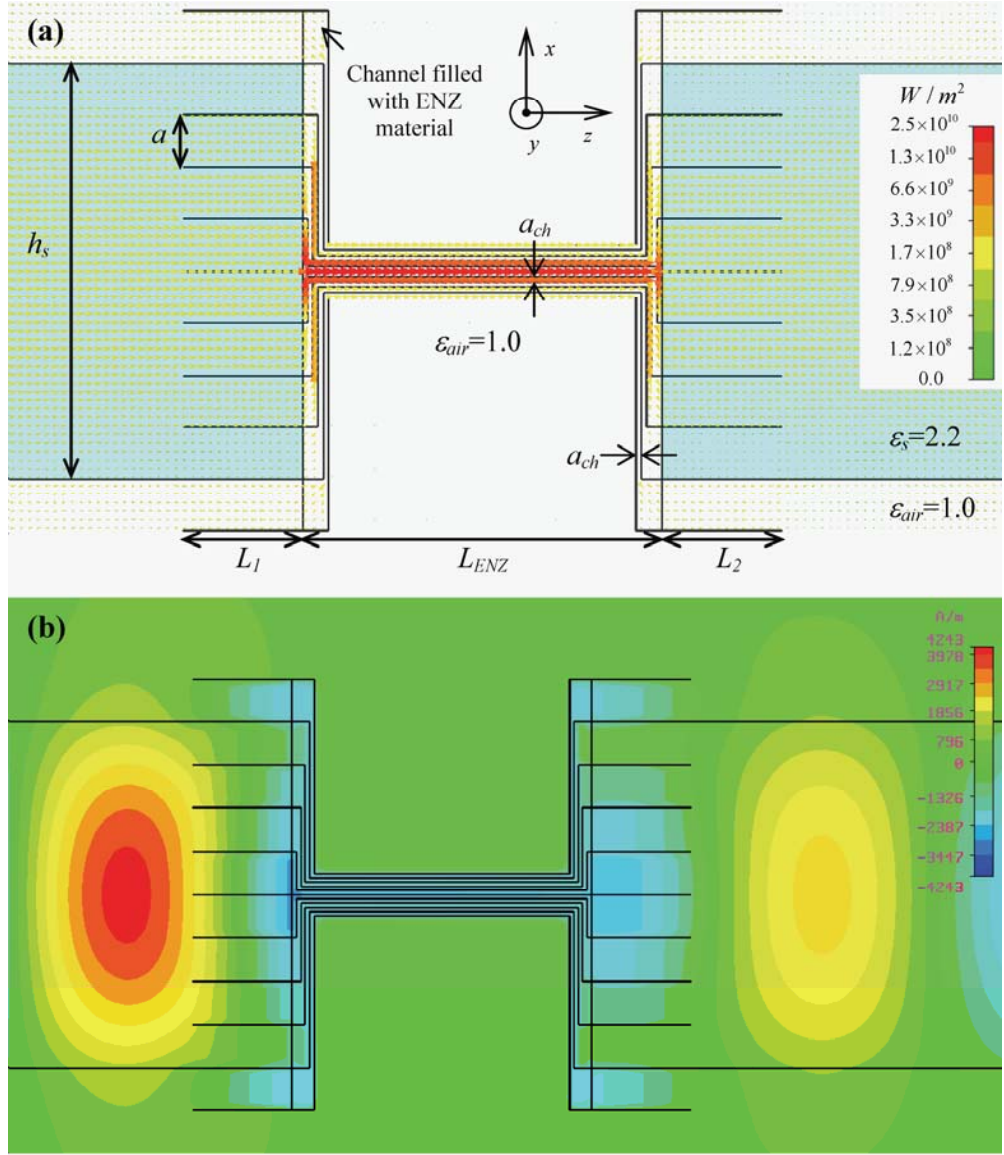

**Fig. 8.** (Color online) Propagation of the fundamental guided mode of a dielectric slab through an array of 10 metallic channels filled with an ENZ material (a) Real part of the Poynting vector. (b) Snapshot (*t=const.*) of the magnetic field. The dimensions of the structure are $L_{ENZ} = 0.75\lambda_d$, $L_1 = L_2 = 0.25\lambda_d$, $a_{ch} = 0.1a$, and $h_s = 0.87\lambda_d$, where $\lambda_d$ is the wavelength in the dielectric. In this example, it was assumed that the permittivity of the ENZ material was $\varepsilon \approx i0.01$.

Hence, in a first step the incoming wave is sampled by an array of parallel-plate waveguides, and each "pixel" of the incoming wave is mapped into the amplitude of the fundamental TEM mode of a specific microstructured waveguide. As illustrated in Fig.

8a, we used 10 metallic waveguides. The spacing between the plates at the input and output planes is $a = h_s/8$. Notice that two of these waveguides sample the fields in the air region. Following the results of section III (see Eq. (6)), we have chosen the length of the metallic plate extensions $L_1$ and $L_2$ such that $L_1 = L_2 = 0.25\lambda_d$. In a second step, the transverse cross-sections of the metallic channels are very much compacted and compressed, as illustrated in Fig. 8a. The distance between the metallic plates is reduced 10 times: $a_{ch} = 0.1a$. To overcome diffraction effects and ensure good transmission through the compacted channels, the channels are filled with an ENZ material. The length of the central channels is $L_{ENZ}$. It is crucial to note that despite the huge shape difference, the propagation characteristics of the channels are rather similar at the design frequency. Indeed, using (7) it is possible to calculate the transmission/reflection coefficient of the TEM mode through each waveguide. Using the fact that $\mu_{r,p} = 1$ (i.e. the ENZ material has no magnetic properties) and $\zeta_{ENZ} = 1 + \rho_{ENZ}$, we obtain that,

$$\zeta_{ENZ} = \frac{2}{2 - i\frac{2\pi}{\lambda_d}\frac{A_p}{a}} \tag{8}$$

The above formula is valid for the metallic waveguides embedded in the dielectric material (a similar formula can be obtained for the two waveguides that stand in the air region). It is simple to estimate the value of $A_p$ (cross sectional area of the ENZ region for a specific waveguide). From Fig. 8a, we see that $A_p \approx (L_{ENZ} + 2na)a_{ch}$, where $n=1$ for the two central waveguides, $n=2$ for the waveguides that lie immediately above or below the two central waveguides, etc. Thus, if (for example) $L_{ENZ} = 1.25\lambda_d$ the transmission coefficient is $\zeta_{ENZ} = 0.91e^{+i25°}$ for the central waveguides ($n=1$), and $\zeta_{ENZ} = 0.83e^{+i34°}$ for the waveguides in the periphery (*n*=4). Thus, it is clear that despite the significant differences in length, the transmission coefficient is relatively similar in all the channels (both in amplitude and phase). For smaller values of $L_{ENZ}$, or for smaller values of $a_{ch}/a$, the transmission coefficients become even more similar (at least provided losses are moderate).

We have used CST Microwave Studio [34] to study the response of the considered structure. To this end we defined two broadband CST waveguide ports that excite the fundamental mode of the dielectric waveguide [37], and defined open boundary conditions at the planes $x = \pm h_s$ (supposing that the dielectric slab is defined by the region $|x| < h_s/2$). In a first example, we consider that $L_{ENZ} = 0.75\lambda_d$ and that $\varepsilon = 0.01i$ (mild losses) at the design frequency. The computed real part of the Poynting vector is depicted in Fig. 8a. It is clearly seen how the real part of Poynting vector lines follow the footprint of the ENZ channels, squeezing the incoming energy through the very tight channels. Notice that from the results of [9] it is expected that the amplitude of the real part of Poynting vector grows by a factor of $a/a_{ch} = 10$ inside the compressed channels. This is completely supported by the full wave simulations depicted in Fig. 8a. We note that the vector Poynting lines are nearly confined inside the dielectric slab or inside the metallic channels, since the amplitude of the fields in the air region is relatively small.

In Fig. 8b, we depict a snapshot of the magnetic field ($H_y$) at some instant $t = const.$ It is seen that $H_y$ is also nearly confined inside the dielectric region or inside the metallic channels. Moreover, consistently with the results of [9], the magnetic field is practically constant inside the ENZ regions. The figure also reveals how the incoming field is

-28-

sampled at the input interface, and then tunneled through the ENZ channels with small amplitude or phase distortions, before it is again coupled to the output dielectric region. Notice that the field distribution at the input and output regions is similar, due to the remarkable imaging properties of the microstructured waveguides.

To further demonstrate the potentials of the described phenomena, we have calculated the magnetic field ($H_y$) at a distance $a/2$ of the $z = const.$ planes that delimit the dielectric regions from the waveguides (from now on, we will refer to these planes as the input and output planes). The plots of the amplitude and phase of $H_y$ are depicted in Fig. 9 for different values of $L_{ENZ}$ and of the ENZ material losses. The remaining parameters are as in the previous example. The solid blue line corresponds to the field at the output plane, the short dashed blue line to the field at the input plane, and the long dashed black line to the field in a uniform dielectric waveguide. To ease the comparison between the different field profiles, all the fields have been normalized so that their amplitude is unity and their phase is shifted to zero at the point $x=0$ of the pertinent $z=const.$ plane (contained in the axis of the dielectric waveguide). We have defined the parameter $T$ as the ratio between the amplitude of the magnetic field at the output plane and the magnetic field at the input plane, both calculated along the axis of the dielectric waveguide ($x=0$). The calculated values of $T$ are shown as insets in Fig. 9.

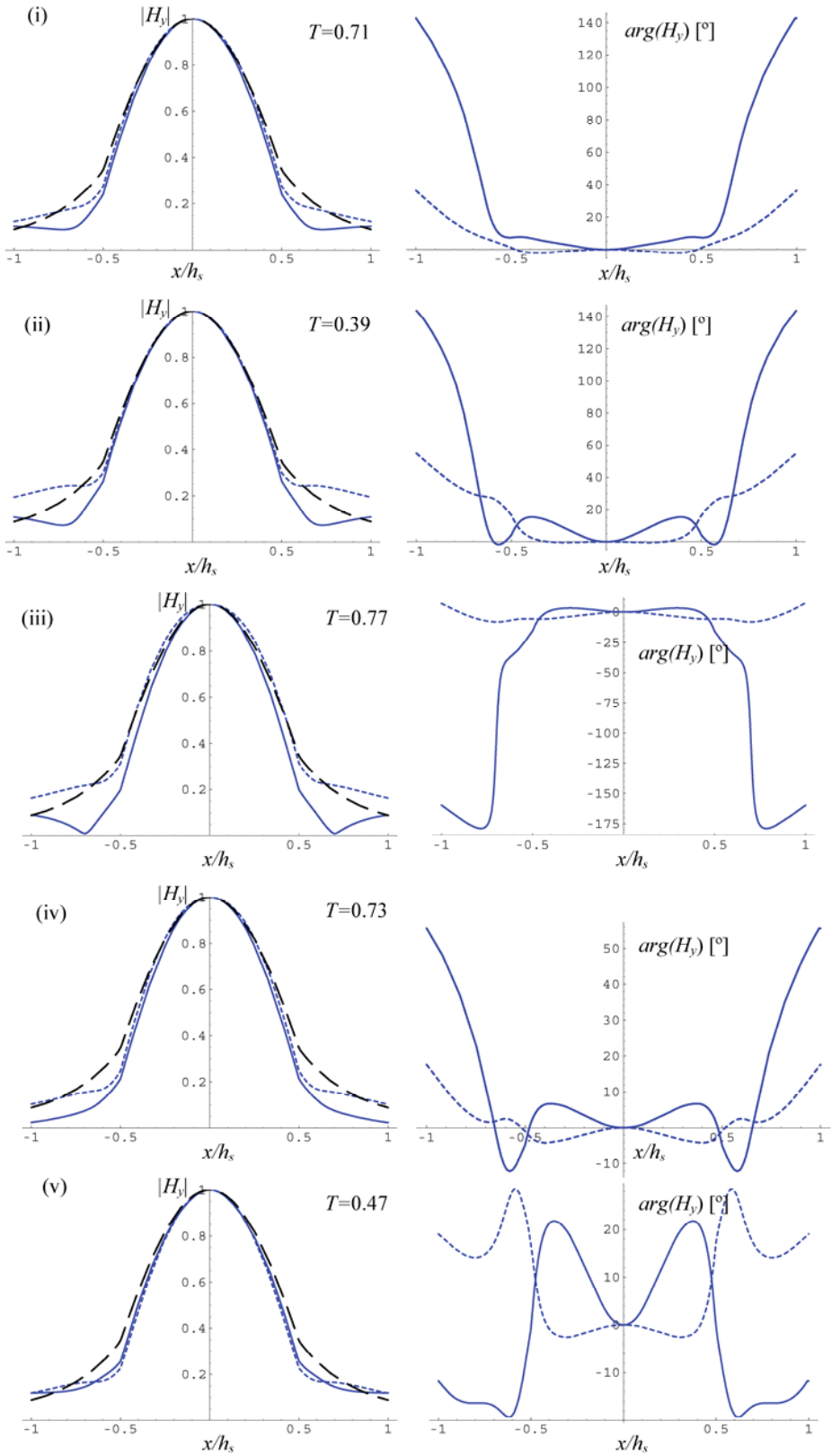

**Fig. 9.** (Color online) Amplitude and phase of the magnetic field at the input plane (short dashed blue line – dark gray in grayscale) and at the output plane (solid blue line – dark gray in grayscale). The fields are normalized so that $H_y = 1$ at the $x=0$ point of the pertinent plane. By definition $T$ (shown as an inset) is the ratio between $H_y$ evaluated at the $x=0$ point of the output and input planes. The long dashed black line represents the profile of the $H_y$ field for a uniform dielectric waveguide. The dimensions of the structure are $L_1 = L_2 = 0.25\lambda_d$, $a_{ch} = 0.1a$, and $h_s = 0.87\lambda_d$. The parameters $L_{ENZ}$ and $\varepsilon$ (permittivity of the ENZ material) are as follows: **(i)** $L_{ENZ} = 1.25\lambda_d$; $\varepsilon \approx i0.001$. **(ii)** $L_{ENZ} = 1.25\lambda_d$; $\varepsilon \approx i0.05$. **(iii)** $L_{ENZ} = 1.00\lambda_d$; $\varepsilon \approx i0.01$. **(iv)** $L_{ENZ} = 0.75\lambda_d$; $\varepsilon \approx i0.01$ **(v)** $L_{ENZ} = 0.5\lambda_d$; $\varepsilon \approx i0.1$ (SiC at $\lambda_0 = 10.31\mu m$).

Note that $T$ is not exactly a transmission coefficient since the wave at the input plane is the superposition of an incident wave and reflected wave. Due to this reason, in case of significant reflections, $T$ may possibly be larger than unity. Notice also that in Fig. 9, the dielectric region corresponds to $|x/h_s| < 0.5$, whereas the air region corresponds to $|x/h_s| > 0.5$. The results depicted in Fig. 9 are quite promising. It can be seen that apart from some attenuation factor ($T < 1$) the amplitude profile of $H_y$ at the output plane is virtually identical to that at the input plane (mainly, in the dielectric region $|x/h_s| < 0.5$). Moreover, the amplitude profile follows very closely that of an uniform dielectric waveguide, with a cosine profile [36, pp. 712] ($H_y\big|_{z=const.} \propto \cos(k_{wg} 2x/h_s)$, where $k_{wg}$ is slightly below $\pi/2$). This demonstrates that the incoming wave is effectively squeezed and guided through the very subwavelength and tight channels, without being perturbed by the microstructured waveguides. Notice that in panel (ii), the length of the central waveguides is as large as $L_{ENZ} = 1.25\lambda_d$ and the losses are as large as $\varepsilon \approx i0.05$, and still a very significant amount of energy is able to reach the output region. A similar comment holds for panel (v), which demonstrates the viability of the proposed effect for an ENZ material readily available in nature (i.e., silicon carbide at $\lambda_0 = 10.31\mu m$ [32]). It is also important to observe that the performance of the proposed "sampling and squeezing" mechanism is to a large extent independent of the specific value of $L_{ENZ}$ (apart from the effect of losses). Indeed, from the results of [9, 10], we expect that if the metallic waveguides filled with the ENZ material are bent or deformed (keeping the area of the corresponding ENZ regions unchanged) the imaging properties of the system will be to a large extent unaffected.

The results for the phase of $H_y$ are also exciting. Indeed, if the incoming wave were not perturbed by the microstructured waveguides the phase of $H_y$ should be constant inside the dielectric waveguide ($|x/h_s| < 0.5$) for a fixed $z = const.$ plane. The results depicted in Fig. 9 demonstrate that such condition is indeed observed to a good approximation. In panel (i) the phase distortion is as small as 15º, even though the difference of paths in the waveguides is a significant fraction of the wavelength ($\lambda_d$). Notice that this value is of the same order of magnitude as that calculated earlier for the phase difference of the transmission coefficients $\zeta_{ENZ}$ associated with the metallic waveguides. Even in case of realistic losses ($\varepsilon \approx i0.1$ in panel (v) for silicon carbide), the phase distortion is only 20º. This suggests very good prospects for the demonstration of the "sampling and squeezing" effect in a realistic setup.

It is interesting to discuss with more detail the possible realization of the setup corresponding to panel (v), in which the ENZ material is SiC ($\lambda_0 = 10.31 \mu m$). In this setup, the distance between the metallic plates is $a = 0.87 \mu m$, and the distance between the compacted waveguides is $a_{ch} = 87nm$. So far, we have assumed that the metals are perfect conductors, and neglected the effect of ohmic losses. As proven in [10, 15], this assumption is justified because the effect of losses in the ENZ material is dominant as compared to the effect of losses in the metals. Moreover, good metals like Ag, Au, and Al have very low losses in the far infrared region [38]. For example, following Ref. [38], in this spectral region aluminum (Al) may be characterized by a Drude-type dispersion model with plasma frequency equal to 3570THz and collision frequency equal to 19.4THz. Thus, the skin depth of aluminum at $\lambda_0 = 10.31 \mu m$ is only $\delta_{Al} = 15 nm$. In particular this shows that it is possible to choose the thickness of the metallic plates, $t_{metal}$, (so far assumed vanishingly thin) such that $\delta_{Al} < t_{metal} \ll a_{ch}$ (as required by the analysis of [10]). Hence, the demonstration of the proposed "sampling and squeezing" effect at $\lambda_0 = 10.31 \mu m$ is indeed within the realm of reality.

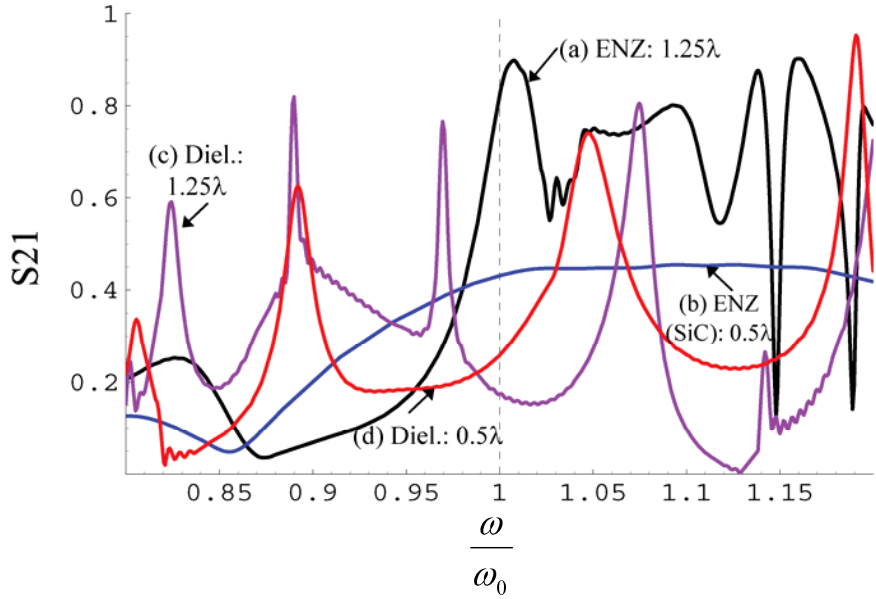

**Fig. 10.** (Color online) $S_{21}$ parameter as a function of the normalized frequency $\omega$. (a) $L_{ENZ} = 1.25 \lambda_d$ and $\varepsilon \approx i0.001$ at $\omega = \omega_0$. (b) $L_{ENZ} = 0.5 \lambda_d$ and $\varepsilon \approx i0.1$ at $\omega = \omega_0$ (SiC at $\lambda_0 = 10.31 \mu m$). (c) $L_{ENZ} = 1.25 \lambda_d$ and ENZ material is replaced by a dielectric with $\varepsilon = 2.2$. (d) $L_{ENZ} = 0.5 \lambda_d$ and ENZ material is replaced by a dielectric with $\varepsilon = 2.2$.

It is also fundamental to have an idea of how variations in frequency may affect the proposed transmission mechanism. To this end, we have computed the $S_{21}$ parameter as a function of $\omega$. The S$_{21}$ parameter (which describes the transmission coefficient for the fundamental dielectric waveguide mode), should not be confused with the parameter $T$ (indicated in the insets of Fig. 9), which defines a "gain" between the near-field distributions at the input and output planes. We have assumed that the ENZ material follows a Drude-type dispersion model with plasma frequency $\omega_p$ and collision frequency $\omega_c$ (as mentioned in section III, this implies that $\varepsilon \approx i\omega_c/\omega_p$ at the plasma frequency; for the case of SiC the assumption of a Drude-type dispersion model around $\omega = \omega_p$ is only a very rough approximation [32], which is nevertheless sufficient for the purposes of this work). The design frequency is $\omega_0 = 2\pi c/\lambda_0$ (note that $\omega_0$ is coincident with $\omega_p$). The calculated S$_{21}$ parameter is depicted in Fig. 10 for different values of $L_{ENZ}$ and $\omega_c/\omega_p$ (curves (a) and (b)). It is seen that S$_{21}$ has a maximum near the design frequency $\omega = \omega_0$. It is also clear that the variation in frequency is relatively smooth around $\omega = \omega_0$, which indicates that the described phenomenon has an acceptable bandwidth. This is particularly true for the case of curve (b), where ENZ losses are significant. We have also calculated the S$_{21}$-characteristic for the remaining examples analyzed in Fig. 9. For the cases (i)-(v) the corresponding S21 is equal to 0.82, 0.33, 0.73, 0.78, and 0.43, respectively, at $\omega = \omega_0$. We note that these values are relatively similar to the corresponding values of $T$ depicted in the insets of Fig. 9.

It is also interesting to study what happens if the ENZ material is replaced by the same material as the dielectric slab. To this end, we have calculated the S$_{21}$-characteristic for the cases $L_{ENZ} = 1.25\lambda_d$ and $L_{ENZ} = 0.5\lambda_d$ (curves (c) and (d) in Fig. 10).

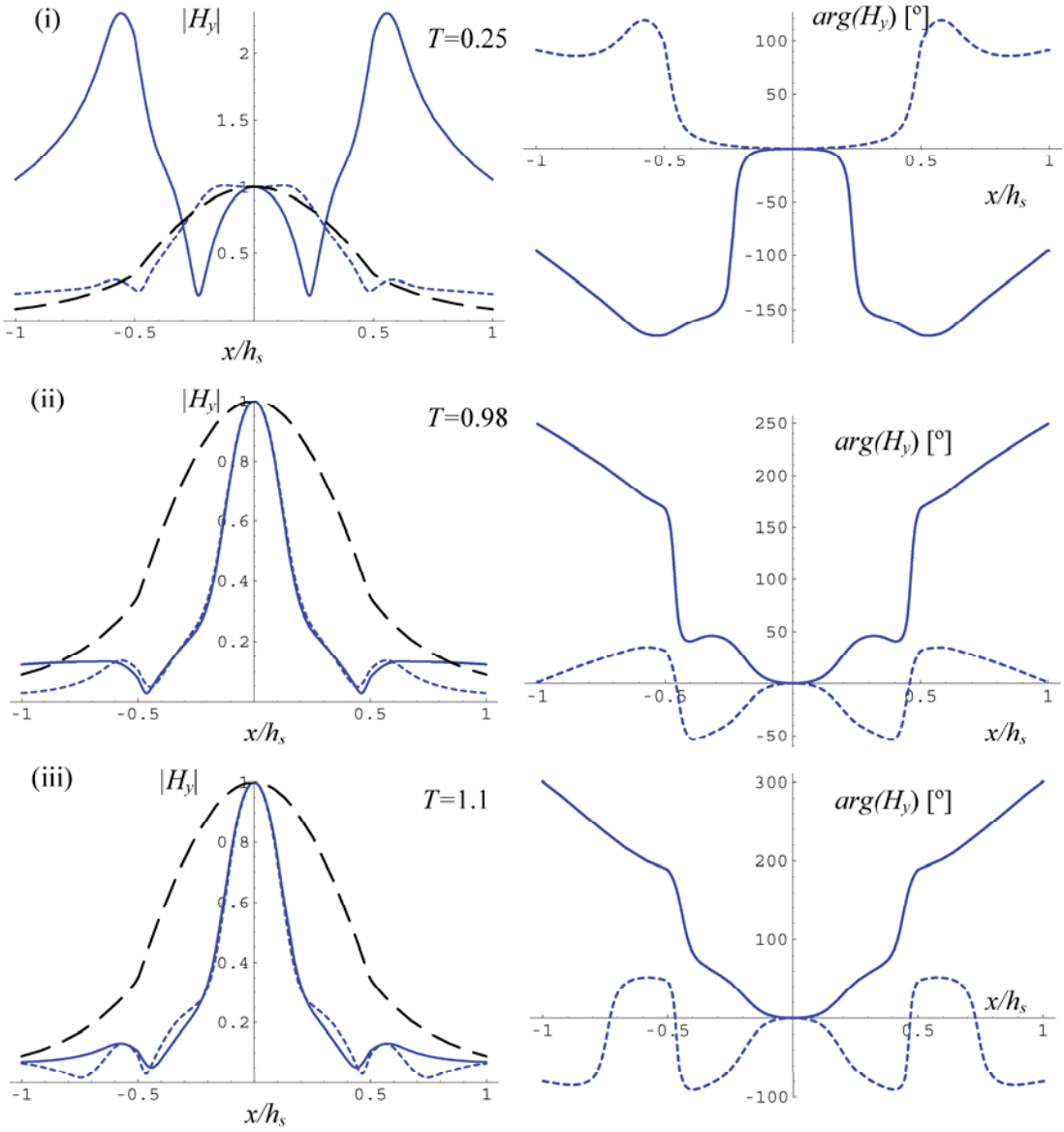

**Fig. 11.** (Color online) Similar to Fig. 9 but the ENZ material is replaced by a dielectric with $\varepsilon = 2.2$. The parameter $L_{ENZ}$ is as follows: **(i)** $L_{ENZ} = 1.25\lambda_d$. **(ii)** $L_{ENZ} = 1.00\lambda_d$. **(iii)** $L_{ENZ} = 0.5\lambda_d$. The fields are evaluated at the frequency **(i)** $\omega = \omega_0$. **(ii)** $\omega = 1.02\omega_0$. **(iii)** $\omega = 1.05\omega_0$, where $\omega_0$ is the design frequency.

Perhaps surprisingly, it is seen that for certain frequencies the $S_{21}$-parameter can be as large as 0.8, even though the metallic waveguides are filled with a material with $\varepsilon = 2.2$. However, as discussed next, the transmission properties in this scenario are fundamentally different from the case where the waveguides are filled with an ENZ material. In order to investigate the main features of this second configuration, we have computed the magnetic field profiles at the input and output planes. The result is depicted in Fig. 11 for several values of $L_{ENZ}$ (now the subscript ENZ has no meaning, and is only kept so that $L_{ENZ}$ can be readily identified in Fig. 8a). We have also calculated the "transmission" factor $T$, which is defined in the same manner as in Fig. 9, and is shown in the insets of Fig. 11. The results of panel (i) (corresponds to the curve (c) in Fig. 10) demonstrate a very significant amplitude distortion at $\omega = \omega_0$ ($|S_{21}| = 0.17$). In fact, the profile of $H_y$ at the output plane is completely different from that at the input plane.

In panel (ii) (corresponds to the curve (d) in Fig. 10), we plot the magnetic field at $\omega = 1.02\omega_0$ (which corresponds to an $S_{21}$ peak with $|S_{21}| = 0.8$). It is seen that even though the profile at the input and output planes is very similar, the field distribution is completely different from that in a uniform dielectric waveguide (long dashed black line). This suggests the emergence of strong reflections next to the waveguides in the periphery of the dielectric waveguide. Notice also that parameter $T$ is very close to unity (even though $|S_{21}| = 0.8$). In fact, as mentioned before, in case of strong reflections it is even possible that $T > 1$. Besides the distortion of amplitude, it is also seen that there is a huge phase variation between the input and output planes (over 100º for $|x/h_s| \leq 0.5$). This shows that when the ENZ material is replaced by the dielectric $\varepsilon = 2.2$ the fields next to the interfaces are strongly perturbed by the metallic waveguides, and that the near field becomes completely corrupted. These properties contrast severely with the results depicted in Fig. 9. The results depicted in panel (iii) are qualitatively similar to those of

-36-

panel (ii). These curves were computed at $\omega = 1.05\omega_0$ in order to hit a peak of the $S_{21}$-parameter ($|S_{21}| = 0.74$). Note that for panel (iii) $T = 1.1$, which as explained before, indicates a strong perturbation of the incoming wave near the input interface.

These results clearly show that metallic waveguides filled with the dielectric material do not, by any means, mimic the behavior of the structure filled with the ENZ material. In fact, the proposed "sampling and squeezing" technique allows transporting an almost *arbitrary* field distribution from the input plane to the output plane, with very subwavelength mode sizes, and negligible amplitude and phase distortions. Such remarkable effect cannot be obtained using a dielectric material.

The peaks of the $S_{21}$ parameter in the case where the metallic plates are filled with a dielectric are clearly due to the emergence of geometrical resonances that enable anomalous transmission through the system (mainly through the central channels). These resonances are strongly dependent on the specific geometry and length of the metallic waveguides. It is also interesting to note the only reason why $S_{21}$ can be relatively large for this configuration is the fact that at the frequency of operation the dielectric waveguide supports a single propagating mode, and thus a few wavelengths away from the output interface (after the near field details are filtered) the narrow profiles shown in Fig. 11 are effectively coupled to the fundamental mode of the dielectric slab. In case of a multimodal waveguide, it is clear that such coupling would be less efficient, and that mode conversion would occur at the output plane. Quite differently, using the "sampling and squeezing" approach such modal conversion would not take place.

To further illustrate the potentials of the "sampling and squeezing" technique and possibility of bending the near-field, we consider the scenario depicted in panel (a) of interface is effectively squeezed, tunneled, and bent through the ultranarrow ENZ channels with zero phase delay. The phase difference between the fields at the input and output interfaces is 180º regardless of the bends of wires with unequal lengths, which corresponds to the propagation phase shift suffered in the parts of the metallic plates embedded in the dielectric.

Fig. 12. It represents two dielectric waveguides with (relative) permittivity $\varepsilon_s = 2.2$ connected at right-angles by a device similar to the one of Fig. 8, formed by an array of 11 metallic plates partially filled with an ENZ material. The thickness of the waveguides $h_s$ is chosen such that the wavelength in the dielectric material is $\lambda_d = 1.15 h_s$, where $\lambda_d = \lambda_0 / \sqrt{\varepsilon_s}$, so that the incoming modal wave is essentially confined to the dielectric slab. As before, the spacing between the metallic plates is $a = h_s / 8$ at the input and output interfaces, and is reduced by a factor of 10 in the narrow channels. The length of the central narrow channel is as large as $L_{ENZ} = 1.2 \lambda_d$. As discussed in section III and Ref. [15], the electric current injected in the ENZ material is tunneled through the ultranarrow channels, being almost insensitive to the bends suffered by the metallic plates, and emerges with nearly the same amplitude and phase at the output (for an ideal lossless $\varepsilon = 0$ limit the amplitude and phase are exactly the same at the input and output). The combined sampling and squeezing mechanisms allow the incoming wave to be effectively transported, rerouted, and bent from one waveguide to the other one, and to compress the modal fields by a factor of 10. Such results are clearly illustrated in panel (b) of Fig. 12, which represents a snapshot in time of the magnetic field distribution, and also in the field animation stored in the supporting online materials [39]. For clarity purposes, and to fully portray the described sampling and squeezing phenomenon, we have assumed in the numerical simulation that the metallic plates are perfect electric conductors and that the ENZ material is characterized by mild losses with $\varepsilon = 0.01i$ at the design frequency. However, similar results are also obtained for materials with a larger loss (e.g. $\varepsilon = 0.1i$), except that the transmitted wave is more attenuated. The field animation [39] clearly shows how the incoming wave after being sampled at the input

To conclude we note that the two-dimensional scenarios studied in the present work can be readily transposed to fully three-dimensional microwave waveguide setups, as in Ref. [11, 15]. Specifically, the footprint of the structure is kept invariant (E-plane) while the device is truncated along the *y*-direction (see Fig. 8), and enclosed in metallic walls. The ENZ response is obtained at microwaves by exploiting the dispersion characteristic of the fundamental $TE_{10}$ waveguide mode in a closed metallic waveguide. More specifically, the H-plane width is tuned so that at the desired frequency of operation the narrow channels (which may be filled with air) effectively behave as if they were filled with an ENZ material. To illustrate these possibilities consider the geometry of Fig. 13a, which shows a modified version of a setup reported in Ref. [15]. As in Ref. [15], the image is produced by an opaque screen with two thin slits (width 0.5*mm*) oriented along the *y*-direction. The opaque screen is illuminated by the fundamental $TE_{10}$ waveguide mode.

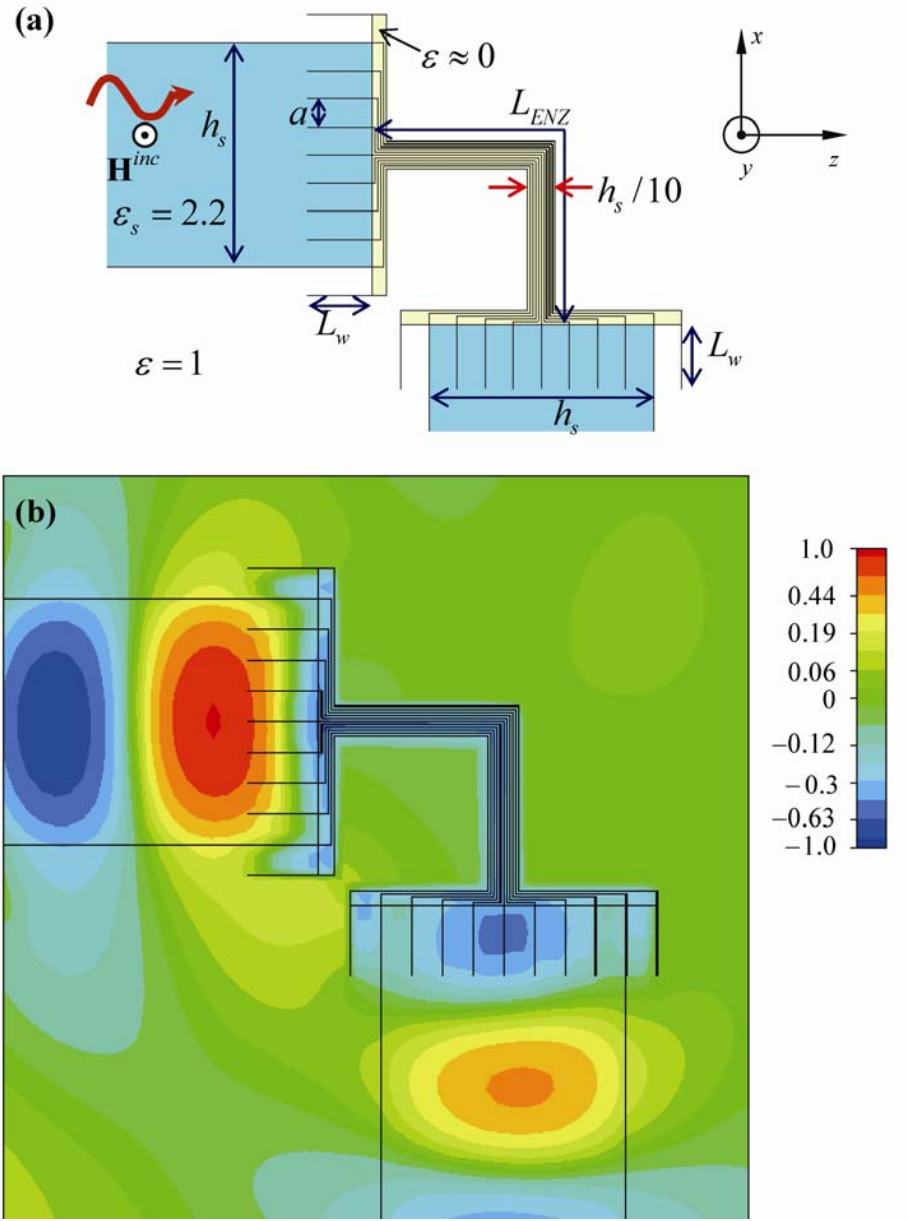

Fig. 12. Sampling, squeezing, and bending the modal field: Panel (a): Two dielectric waveguides are connected at right angles by an array of 11 metallic plates partially embedded in an ENZ material. The incoming surface wave mode is sampled by the array of metallic waveguides and is squeezed, compressed, and bent through ultranarrow parallel-plate channels with a 90º-bend. Panel (b): Snapshot of the magnetic field at $t$=0 (normalized to arbitrary unities). Notice how the magnetic field is effectively frozen and tunneled through the ENZ channels with zero phase-shift. The phase shift with respect to the input and output interfaces is approximately 180º for every "pixel".

As in Ref. [11], the ultranarrow channels are filled with air while the wide channels are filled with Teflon ($\varepsilon_d = 2.0$). The H-plane width, $W_H = 102 mm$, is tuned so that at the frequency of operation ($1.47 \, GHz$) the regions filled with Teflon behave effectively as a continuous material with $\varepsilon_{eff} = 1.0$, whereas the narrow channels filled with air behave as a material with $\varepsilon_{eff} = 0$. The remaining dimensions of the structure are $L_w = 0.25 \lambda_0 / \sqrt{\varepsilon_{eff,Teflon}} = 51 mm$, $a = W_E / 10 = 5.6 mm$, $L_{ENZ} = 102 mm$, and the waveguide walls are assumed to be made of copper ($\sigma = 5.8 \times 10^7 \, S/m$). The distance between the two slits at the input plane is $33.6 mm$. In Fig. 13b and Fig. 13c, we report a snapshot of the $E_z$ component of the electric field, at the input and output plane, respectively. Consistent with the results of Ref. [15], it is seen that notwithstanding with the complex manipulations performed in the image, and with the fact that the propagation channel suffers a 10-fold compression and the narrow channels are operated at cut-off, the image is effectively transported by the proposed device from the input to the output plane, preserving the near-field information.

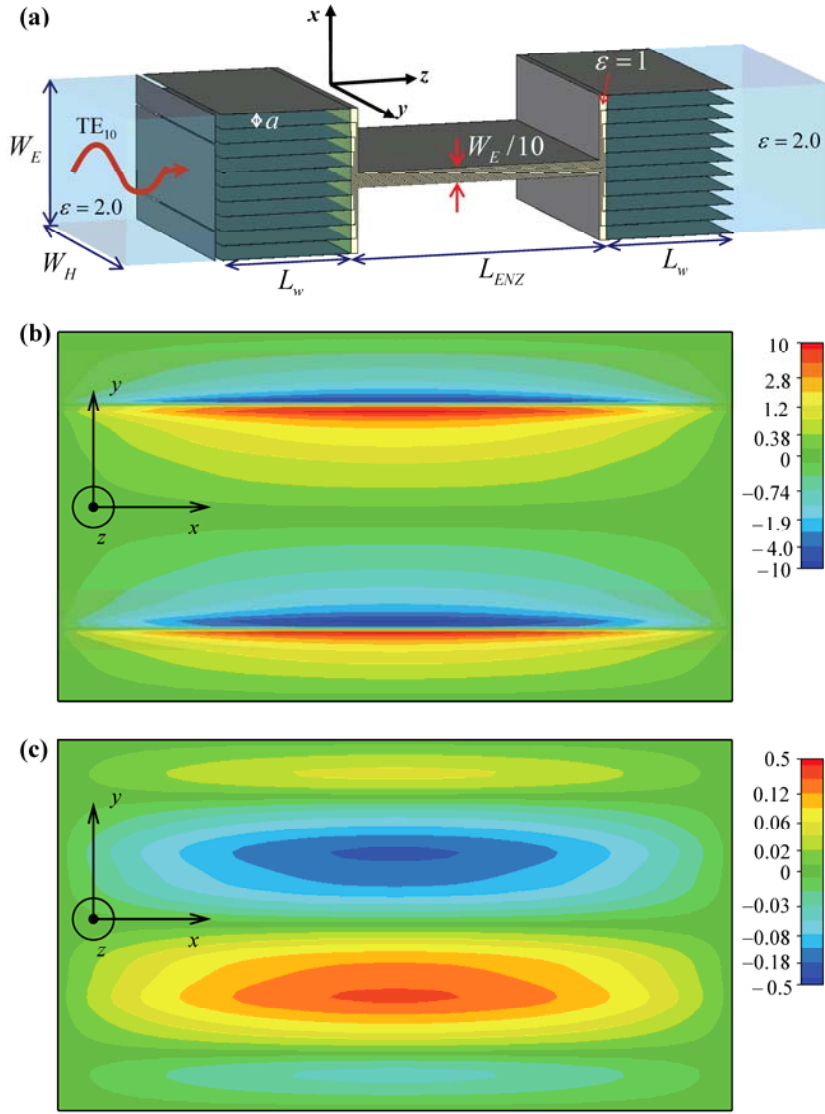

Fig. 13. "Sampling and squeezing" with a microwave waveguide: Panel (a): Two microwave rectangular metallic waveguides filled with Teflon ($\varepsilon = 2.0$) are connected by an array of ultranarrow channels filled with air ($\varepsilon = 1.0$). Panel (b): Snapshot (*t*=0) of the electric field normal component, $E_z$, (normalized to arbitrary unities) at the input plane. Panel (c): Similar to panel (b) but referred to the output plane.

# V. Conclusions

In this work, we studied in detail the physical mechanisms that enable the "sampling and squeezing" effect in scenarios where the fields are compressed along a single direction of space. It was shown that an array of metallic parallel plates may convert an incoming wave into a collection of TEM modes, which may then be squeezed through very tight arbitrarily shaped channels filled with an ENZ material, with little amplitude or phase distortion. We derived simple theoretical models that enable the description of these phenomena using analytical methods. The "sampling and squeezing" concepts may allow guiding a wave with mode sizes well below the diffraction limit, or to compress a complex image several folds without resolution loss. Our numerical simulations suggest that silicon carbide (SiC) may be used to demonstrate experimentally the described phenomena at $\lambda_0 = 10.31 \mu m$. These results may have interesting potentials in imaging and sensing applications in the nanoscale.

## Acknowledgments:

This work is supported in part by the U.S. Office of Naval Research (ONR) Multidisciplinary University Research Initiative (MURI) grant number N00014-10-1-0942 and by Fundação para Ciência e a Tecnologia under project PTDC/EEA-TEL/100245/2008.

# Appendix A

Here we demonstrate that provided the lattice constant *a* is considerably smaller than the wavelength of light, the array of parallel metallic plates may be regarded as an artificial material.

In Ref. [25], we have extended the concepts proposed in [40], and demonstrated that under some conditions a metamaterial formed by metallic plates may effectively behave as a continuous material with negative permittivity. We have derived a simple model that allows taking into account interface effects, and explained how to use these metamaterials in the design of cloaking devices that may significantly reduce the total scattering cross-section of a given obstacle. However, the model described in [25, 40] assumes that the electric field is polarized along the *y*-direction. This is not the case in this work, since here the electric field is in the *x*-*z* plane as illustrated in Fig. 1a. As described below, the electric properties of the material for this polarization are completely different from those described in [25, 40].

Indeed, let us suppose that the parallel-plate medium can be described by a diagonal (relative) permittivity dyadic with components $\varepsilon_{xx}$ and $\varepsilon_{zz}$. Thus, an H-polarized wave (i.e. with magnetic field along *y*) and propagation factor $e^{+i\mathbf{k}\cdot\mathbf{r}}$ ($\mathbf{k} = (k_x, 0, k_z)$ is the wave vector) must satisfy the dispersion relation:

$$\frac{k_x^2}{\varepsilon_{zz}} + \frac{k_z^2}{\varepsilon_{xx}} = \left(\frac{\omega}{c}\right)^2 \tag{A1}$$

From Fig. 1a, it is obvious that $\varepsilon_{xx} = 1$. Thus, solving the above equation for $k_z$, and imposing that the obtained function, $k_z = k_z(\omega, k_x)$, is coincident with the dispersion relation of the fundamental TEM waveguide mode, $k_z = i\gamma_0 = \omega/c$ we find that $\varepsilon_{zz}$ must be such that $\varepsilon_{zz} = \infty$. Thus, this simple analysis suggests that for H-polarization, the fundamental mode of the artificial material may be described by the effective parameters:

$$\varepsilon_{xx} = 1, \qquad \varepsilon_{zz} = \infty \qquad \text{(fundamental TEM mode)} \qquad (A2)$$

This suggests that the parallel plate medium may behave as a material with extreme optical anisotropy (a medium with such characteristics was designated as "wave-guiding medium" in [23]).

It is important to emphasize that the derived effective parameters only apply to the fundamental TEM mode. Higher order modes cannot be described by the same model. Thus, the artificial medium is characterized by spatial dispersion effects [41] (which ultimately is the case of all composite materials). In what follows, we prove that it is possible to consider an alternative and more accurate spatially dispersive model for the effective permittivity of the structured material, and explain how this model can be used to predict the transmission properties of a material slab and calculate the corresponding electromagnetic transfer function.

The objective is to obtain a spatially dispersive dielectric function that may describe not only the behavior of the fundamental waveguide mode (with $m=0$), but also the behavior of the next mode (with $m=1$). We still require that $\varepsilon_{xx} = 1$ (since the polarizability of the inclusions along the $x$-direction is necessarily zero), and try to find an adequate $\varepsilon_{zz} = \varepsilon_{zz}(\omega, \mathbf{k})$ that characterizes the composite medium.

Using $\varepsilon_{xx} = 1$ and Eq. (A1), it follows that the dispersion characteristics of the plane wave solutions in the metamaterial can be rewritten as:

$$\varepsilon_{zz} = 1 - \frac{\left(\dfrac{\omega}{c}\right)^2 - k_z^2 - k_x^2}{\left(\dfrac{\omega}{c}\right)^2 - k_z^2} \tag{A3}$$

Next, we note that the relation $\left(\dfrac{\omega}{c}\right)^2 - k_z^2 = \left(m\dfrac{\pi}{a}\right)^2$ is satisfied by the *m*-th waveguide mode (with propagation constant $\gamma_m$). In particular, the fundamental TEM mode corresponds to the case *m*=0. Because the TEM mode satisfies $k_z^2 = \left(\dfrac{\omega}{c}\right)^2$ it is clear that (A3) implies that $\varepsilon_{zz} = \infty$, consistently with the discussion in the beginning of this section. Since $\left(\dfrac{\omega}{c}\right)^2 - k_z^2 = \left(\dfrac{\pi}{a}\right)^2$ should be verified for the *m*=1 mode, we replace $\left(\dfrac{\omega}{c}\right)^2 - k_z^2$ by $\left(\dfrac{\pi}{a}\right)^2$ in the numerator of the fraction in the right-hand side of (A3) (the denominator of (A3) is kept unchanged because otherwise the TEM mode properties would not be correctly predicted). In this way, we obtain the following dielectric function,

$$\varepsilon_{xx} = 1, \qquad \varepsilon_{zz} = 1 - \frac{\left(\dfrac{\omega_p}{c}\right)^2 - k_x^2}{\left(\dfrac{\omega}{c}\right)^2 - k_z^2} \tag{A4}$$

where $\omega_p = \dfrac{\pi}{a}c$ is by definition the plasma frequency of the parallel-plate medium. This dielectric function, constructed by hand, fully describes the properties of the modes with *m*=0 and *m*=1. In fact, if we substitute (A4) into (A1) and solve the corresponding equation with respect to $\omega$, we easily find that there are exactly two possible solutions: $\omega^2 = c^2 k_z^2$ or $\omega^2 = \omega_p^2 + c^2 k_z^2$. These are precisely the dispersion characteristics of the waveguide modes with *m*=0 and *m*=1. Note that from (A4) it is evident that for long wavelengths the mode with *m*=1 sees effectively a medium with negative permittivity. Equation (A4) also shows that the macroscopic electric field associated with a plane wave of the form $\mathbf{H} = H_0 e^{i(k_x x + k_z z)} \hat{\mathbf{u}}_y$ is given by:

$$\mathbf{E} = \frac{1}{\omega \varepsilon_0} \left( \frac{k_z}{\varepsilon_{xx}} \hat{\mathbf{u}}_x - \frac{k_x}{\varepsilon_{zz}(\omega, \mathbf{k})} \hat{\mathbf{u}}_z \right) H_0 e^{i(k_x x + k_z z)} \tag{A5}$$

In some sense the proposed permittivity model does not bring much into the table, since we already knew all the properties of the modal solutions supported by the parallel plate medium, and Eq. (A4) can hardly be used to make new predictions. However, Eq. (A4) has a salient feature: it is the direct analog of the permittivity model proposed in Ref. [42] to characterize the electromagnetic properties of a wire medium. In fact, it can be checked that $\varepsilon_{zz}$ given by Eq. (A4) is very similar to the corresponding formula of [42]. The only difference is that $\varepsilon_{zz}$ given by (A4) depends on both $k_x$ and $k_z$, while in the wire medium model $\varepsilon_{zz}$ only depends on $k_z$. The physical reason for this property is evident: unlike the wire medium, the air region in the parallel-plate medium is not connected (for infinitely extended metallic plates). Thus, the spatial dispersion effects are more severe in the parallel-plate medium than in the wire medium. More than a curiosity, this analogy between the parallel-plate medium and the wire medium turns out to be very useful, because it tells us how to characterize the transmission of electromagnetic waves through such metamaterial using homogenization techniques. Before describing these results, it is important to stress that (A4) assumes that the fields are H-polarized and the wave propagates in the *x-z* plane. The formula can be extended to a general case, but such discussion is outside the scope of this work.

Let us now consider again the scattering problem depicted in Fig. 1a. We will explain how such problem can be conveniently solved using homogenization techniques. To this end, we replace the periodic array of parallel plates by a homogenized slab characterized by the nonlocal dielectric function (A4). As discussed above, such model predicts that two different H-polarized waves may propagate in the artificial material. Thus, the magnetic field is given by:

$$H_y = H_0 e^{+ik_x x} \times \begin{cases} e^{-\Gamma_0 z} + R e^{+\Gamma_0 z}, & z < 0 \\ A_+ e^{-\gamma_0 z} + A_- e^{+\gamma_0 z} + B_+ e^{-\gamma_1 z} + B_- e^{+\gamma_1 z}, & 0 < z < L \\ T e^{-\Gamma_0 (z-L)}, & z > L \end{cases} \quad (A6)$$

where $R$ is the reflection coefficient, $T$ is the transmission coefficient, and $A_\pm$ and $B_\pm$ are the coefficients associated with the modes inside the homogenized slab. The corresponding formula for the macroscopic electric field can be easily obtained with the help of (A5), noting that $H_y$ is a superposition of plane waves.

In order to calculate the unknown coefficients ($R$, $T$, $A_\pm$ and $B_\pm$), we need to impose boundary conditions at the interfaces between the parallel-plate medium slab and the air regions, namely we need to impose the continuity of the tangential components of the electromagnetic fields ($H_y$ and $E_x$) at the interfaces. However, it is simple to verify that such boundary conditions are insufficient to calculate the unknown parameters. This is a consequence of the spatial dispersion of the homogenized slab. A similar discussion can be found in Refs. [18, 27, 43]. In Ref. [43] it was demonstrated that an additional boundary condition (ABC) is required to solve unambiguously a related scattering problem for a wire medium. The ABC proposed in [43] imposes the continuity of the normal component of the electric field $E_z$ (assuming that the host of the composite material is air). As proven in Ref. [43], such ABC is a direct consequence of the fact that the microscopic electric current that flows along the metallic inclusions must vanish at the interfaces. Thus, exploring the direct analogy between the structured material studied in this work and the wire medium, we conclude that to determine the unknown coefficients in (A6) we need to impose not only the continuity of $H_y$ and $E_x$, but also the continuity of $E_z$. Following [18, 43], we may easily prove that such boundary conditions are equivalent to (notice that in this work we assume that the metallic plates are filled with a material with the same electric properties as the air regions):

$$\left[H_y\right]=0; \qquad \left[\frac{dH_y}{dz}\right]=0; \qquad \left[\frac{d^2H_y}{dz^2}\right]=0 \qquad (A7)$$

where $[...]$ represents the jump discontinuity of the function inside rectangular brackets at the interface, i.e. the function evaluated at the air side minus the function evaluated at the artificial medium side. In this way, proceeding as in [27, 43] and imposing the boundary conditions given by (A7) at the interfaces $z=0$ and $z=L$, we easily obtain the 6×6 linear system:

$$\begin{pmatrix} -1 & 1 & 1 & 1 & 1 & 0 \\ -\Gamma_0 & -\gamma_0 & \gamma_0 & -\gamma_1 & \gamma_1 & 0 \\ -\Gamma_0^2 & \gamma_0^2 & \gamma_0^2 & \gamma_1^2 & \gamma_1^2 & 0 \\ 0 & e^{-\gamma_0 L} & e^{+\gamma_0 L} & e^{-\gamma_1 L} & e^{+\gamma_1 L} & -1 \\ 0 & -\gamma_0 e^{-\gamma_0 L} & \gamma_0 e^{+\gamma_0 L} & -\gamma_1 e^{-\gamma_1 L} & \gamma_1 e^{+\gamma_1 L} & \Gamma_0 \\ 0 & \gamma_0^2 e^{-\gamma_0 L} & \gamma_0^2 e^{+\gamma_0 L} & \gamma_1^2 e^{-\gamma_1 L} & \gamma_1^2 e^{+\gamma_1 L} & -\Gamma_0^2 \end{pmatrix} \begin{pmatrix} R \\ A_+ \\ A_- \\ B_+ \\ B_- \\ T \end{pmatrix} = \begin{pmatrix} 1 \\ -\Gamma_0 \\ \Gamma_0^2 \\ 0 \\ 0 \\ 0 \end{pmatrix} \qquad (A8)$$

This system can be easily solved either numerically or even analytically (see Refs. [18, 27] for a related result).

At this point, we note that (A6) is essentially equivalent to assuming that the waveguide modes with $m \geq 2$ are not significantly excited, and that the reflected and transmitted waves consist only of the fundamental Floquet mode (compare with Eqs. (1) and (2)). Even though this assumption is fairly accurate for long wavelengths, the homogenization technique can still be further refined to take into account the effect of higher order diffraction modes. In fact, in [25] we studied a canonical reflection problem (scattering of plane waves by a semi-infinite slab of the parallel-plate medium), and theoretically demonstrated that in order to properly homogenize the artificial material and take into account the fluctuations of the fields near the interfaces with air, it is necessary to introduce virtual interfaces displaced from a distance $\delta$ from the actual physical interfaces. We proved that a slab of the parallel-plate material with thickness $L$ is effectively equivalent to a homogenized slab with thickness $L_{ef} = L + 2\delta$. Even though the results derived in [25] assume that the fields are E-polarized (electric field is along the *y*-direction), it can be easily checked that the formula obtained in [25] for $\delta$ is still valid for the H-polarization case considered here, provided one uses the spatially dispersive model implicit in (A6) (see also Ref. [43]). Thus, from [25] we conclude that $\delta$ is given by Eq. (4).

In practice, to take into account the effect of virtual interfaces we may proceed as follows. Suppose that the metallic plates are extended from $z_0 < z < z_0 + L$, where $z = z_0$ and $z = z_0 + L$ define the physical interfaces with air. Then, the theory derived in [25] implies that this periodic structure is equivalent to a homogenized slab extended from $z_0 - \delta < z < z_0 + L + \delta$ (i.e. with thickness $L_{ef} = L + 2\delta$), and (virtual) interfaces at $z = z_0 - \delta$ and $z = z_0 + L + \delta$. Thus, it is expected that the scattering parameters calculated at the virtual interfaces using the dielectric function (A6) are nearly coincident with the *exact* scattering parameters (obtained for the periodic array of metallic plates) calculated at the same reference planes. The results reported in section II completely support these claims.

# Appendix B

Here, we derive a simple analytical method to characterize the propagation of H-polarized electromagnetic waves through the system depicted in Fig. 1b. To this end, we will generalize the homogenization technique developed in Appendix A. For simplicity, we assume that the permittivity of the ENZ material is exactly $\varepsilon = 0$, even though this assumption is not really essential. The incident wave has magnetic field $H_y = H_0 e^{+ik_x x} e^{-\Gamma_0 z}$. Following the results of Appendix A, we will assume that the wave inside the metallic waveguides can be written only in terms of the fundamental TEM mode (with propagation constant $\gamma_0 = -i\omega/c$) and of the $m=1$ mode (with propagation constant $\gamma_1$). Thus, for $0 < z < L_1$ (the left-hand side region of the metallic waveguides filled with air) the magnetic field is to a first approximation,

$$H_y = H_0 e^{+ik_x x}\left( A_1^+ e^{-\gamma_0 z} + A_1^- e^{+\gamma_0 z} + B_1^+ e^{-\gamma_1 z} + B_1^- e^{+\gamma_1 z} \right), \qquad 0 < z < L_1 \tag{B1}$$

where $A_1^\pm$ and $B_1^\pm$ are the unknown coefficients of the expansion. Next, we use the fact that for oblique incidence the ENZ slab behaves essentially as perfect magnetic conductor (PMC) with reflection coefficient (for the magnetic field) equal to $-1$. Thus, the waveguide mode associated with $m=1$ is completely reflected at the interface $z = L_1$. Hence, in the ENZ limit Eq. (B1) can be rewritten as,

$$H_y = H_0 e^{+ik_x x}\left( A_1^+ e^{-\gamma_0 z} + A_1^- e^{+\gamma_0 z} + B_1 \frac{\sinh(\gamma_1(z-L_1))}{-\sinh(\gamma_1 L_1)} \right), \qquad 0 < z < L_1 \tag{B2}$$

where $B_1$ is some unknown constant. Similarly, for $L_1 + L_{ENZ} < z < L_{tot}$, we have that:

$$H_y = H_0 e^{+ik_x x} \left( A_2^+ e^{-\gamma_0(z-L_{tot})} + A_2^- e^{+\gamma_0(z-L_{tot})} + B_2 \frac{\sinh(\gamma_1(z-L_t+L_2))}{\sinh(\gamma_1 L_2)} \right),$$
$$L_1 + L_{ENZ} < z < L_{tot} \quad (B3)$$

where $L_{tot} = L_1 + L_{ENZ} + L_2$, and $A_2^\pm$ and $B_2$ are unknown coefficients.

On the other hand, in the air regions, we neglect all the Floquet modes, except for the fundamental mode. Hence, as in Appendix A, we may write that:

$$H_y = H_0 e^{+ik_x x} \times \begin{cases} e^{-\Gamma_0 z} + Re^{+\Gamma_0 z}, & z < 0 \\ Te^{-\Gamma_0(z-L_{tot})}, & z > L_{tot} \end{cases} \quad (B4)$$

where $R$ and $T$ are the reflection and transmission coefficients, respectively.

It is interesting to note that there is a simple relation between the coefficients $A_1^\pm$ and the coefficients $A_2^\pm$ in the ENZ limit. In fact, consider a generic waveguide and let $\rho_{ENZ}$ and $\zeta_{ENZ}$ be the reflection and transmission coefficients for the magnetic field, respectively, when the fundamental TEM mode impinges on the ENZ transition (in this definition, it is implicit that the input and output sections of the waveguide are infinitely extended as in [9, 10]). In the ENZ limit, we have that $\zeta_{ENZ} = 1 + \rho_{ENZ}$ [9, 10]. Hence, it is straightforward to verify that for the configuration of Fig. 1b, we have that:

$$A_1^- = A_1^+ e^{-2\gamma_0 L_1} \rho_{ENZ} + A_2^- e^{-\gamma_0(L_1+L_2)}(1+\rho_{ENZ}) \quad (B5a)$$
$$A_2^+ = A_1^+ e^{-\gamma_0(L_1+L_2)}(1+\rho_{ENZ}) + A_2^- e^{-2\gamma_0 L_2} \rho_{ENZ} \quad (B5b)$$

It is simple to check that the parameter $\rho_{ENZ}$ is given by:

$$\rho_{ENZ} = \frac{+i\dfrac{\omega}{c}\mu_{r,p}\dfrac{A_p}{a}}{2-i\dfrac{\omega}{c}\mu_{r,p}\dfrac{A_p}{a}} \quad \text{(ENZ lossless limit)} \quad (B5c)$$

where $\mu_{r,p}$ is the relative permeability of the ENZ material, $a$ is the spacing between the metallic plates, and $A_p = aL_{ENZ}$ (see Fig. 1b).

To calculate the reflection and transmission coefficients $R$ and $T$, and the remaining unknown coefficients $A_1^\pm$, $A_2^\pm$, $B_1$, and $B_2$, we need to impose boundary conditions at the interfaces with the air region $z = 0$ and $z = L_{tot}$. Since the continuity of the tangential electric field at the interfaces with air is equivalent to the continuity of $dH_y/dz$ (see Appendix A), we find that the continuity of $H_y$ and $E_x$ yield the following equations:

$$1 + R = A_1^+ + A_1^- + B_1 \tag{B5d}$$
$$(1 - R)(-\Gamma_0) = (A_1^+ - A_1^-)(-\gamma_0) - B_1 \coth(\gamma_1 L_1)\gamma_1 \tag{B5e}$$
$$T = A_2^+ + A_2^- + B_2 \tag{B5f}$$
$$T(-\Gamma_0) = (A_2^+ - A_2^-)(-\gamma_0) + B_2 \coth(\gamma_1 L_2)\gamma_1 \tag{B5g}$$

The system formed by equations (B5a)-(B5b) and (B5d)-(B5g) is underdetermined, because we have 8 unknowns and only 6 equations. However, noting that the interaction of the set of parallel-plates with the air regions can obviously be studied using the same methods as in Appendix A, it is clear that as in Appendix A we need to consider an additional boundary condition (ABC) to remove the extra degrees of freedom of the problem. As discussed in Appendix A, this boundary condition is the continuity of $E_z$ at the interfaces with air, and is equivalent to the continuity of $d^2 H_y / dz^2$. Hence, we obtain the equations:

$$(1+R)\Gamma_0^2 = (A_1^+ + A_1^-)\gamma_0^2 + B_1 \gamma_1^2 \tag{B5h}$$
$$T\Gamma_0^2 = (A_2^+ + A_2^-)\gamma_0^2 + B_2 \gamma_1^2 \tag{B5i}$$

By solving numerically the 8×8 linear system formed by (B5a)-(B5b) and (B5d)-(B5i), we may now easily obtain $R$ and $T$ using the proposed analytical formalism.

As in Appendix A, it is possible to improve the accuracy of the proposed homogenization method by introducing virtual interfaces displaced from the physical interfaces a distance $\delta = 0.1a$. This means that the effective length of the parallel-plate extensions is given by $L_{1ef} = L_1 + \delta$ and $L_{2ef} = L_2 + \delta$. Thus, when computing the scattering parameters using the proposed analytical formalism, it is recommended that one replaces $L_1$ and $L_2$ by $L_{1ef}$ and $L_{2ef}$, respectively, in system (B5).